\documentclass[12pt]{article}
\usepackage{natbib}
\usepackage{amsmath}
\usepackage{url}
\usepackage{pgf, tikz}
\usepackage{float}
\usepackage{parskip}

\usepackage{etoolbox}
\makeatletter
\patchcmd{\@makecaption}
  {\parbox}
  {\advance\@tempdima-\fontdimen2} 
  {}{}
\makeatother  

\usepackage[top=0.7in, bottom=0.7in, left=1in, right=1in]{geometry}

\begin{document}
\begin{center}
{\LARGE Incorporating testing volume into  estimation of effective reproduction number dynamics}\\
\vspace{0.2cm}

Isaac H. Goldstein$^{1}$, Jon Wakefield$^{2}$, and Volodymyr M. Minin$^{1,*}$\\
\vspace{0.2cm}

$^{1}$Department of Statistics, University of California, Irvine\\
$^{2}$Departments of Biostatistics and Statistics, University of Washington,
Seattle\\
$^{*}$\url{vminin@uci.edu}

\end{center}

\begin{abstract}
Branching process inspired models are widely used to estimate the effective reproduction number --- a useful summary statistic describing an infectious disease outbreak --- using counts of new cases. Case data is a real-time indicator of changes in the reproduction number, but is challenging to work with because cases fluctuate due to factors unrelated to the number of new infections. We develop a new model that incorporates the number of diagnostic tests as a surveillance model covariate. Using simulated data and data from the SARS-CoV-2 pandemic in California, we demonstrate that incorporating tests leads to improved performance over the state-of-the-art.
\end{abstract}
\section{Introduction}
In an infectious disease epidemic, the effective reproduction number is the average number of people a newly infected person will subsequently infect. 
When the effective reproduction number is above one, an epidemic is out of control and will continue to grow, vice versa if it is below one. 
This makes the effective reproduction number a useful summary of the state of an epidemic which can provide guidance to policy makers. 
As such, estimates of the effective reproduction number based on observed data can be an important part of any public health response during an epidemic. 
Recent examples from the SARS-CoV-2 pandemic include work by \citet{Mishra2020} in Scotland, as well as efforts by \citet{swiss_taskforce}. 
\par 
An early effort of using a likelihood based approach to estimate the effective reproduction number is that of \citet{wallinga2004different}, which is based on modeling transmission trees.
A recently popular class of estimators for the effective reproduction number (used in both \citep{Mishra2020} and \citep{swiss_taskforce}) is inspired by stochastic branching process models, where infectious individuals infect a random number of new individuals at random points in time.
The most widely used model in this class is available in the \texttt{EpiEstim} R package \citep{cori_new_2013,thompson2019improved}, which is based on ideas put forth by \citet{Fraser2007}.
\texttt{EpiEstim} assumes all new infections (incidence) are observed, and uses a time series of observed cases as data. 
During the SARS-CoV-2 pandemic, a number of methods in this class of estimators have been developed.
The methods of \citet{parag2021improved} and \citet{capistran2022filtering} continue to assume incidence (or incidence up to a constant) are observed, and focus on improving how changes in $R_{t}$ are modeled over time, while avoiding Markov chain Monte Carlo based methodologies.
The methods of \citet{abbott_estimating_2020}, \citet{huisman2022estimation}, \citet{epidemia}, and \citet{epidemia_paper} use more computationally intensive approaches which model observed data as functions of latent incidence, either through explicit Bayesian models \citep{abbott_estimating_2020,epidemia,epidemia_paper} or through a pipeline that first bootstraps latent incidence which is then used as input into \texttt{EpiEstim}. 
The methods of \citet{teh2022efficient}, \citet{epidemia}, and \cite{epidemia_paper} also begin to tackle the problem of how to estimate $R_{t}$ across spatial locations.
Many of these methods have not been scrutinized via extensive simulation studies under model mis-specification, or in some cases, not probed at all, making it difficult to understand the strengths and weaknesses  of this class of methods.
\par 
This gap in knowledge is particularly relevant when it comes to applying such methods to observed case counts of an infectious disease. 
As the SARS-CoV-2 pandemic has demonstrated, observed cases of an infectious disease are often circuitously related to the true number of new infections, due to constraints in testing supply, asymptomatic infections, testing eligibility, and reporting delays. 
These factors can make estimating the effective reproduction number from cases quite difficult in real world situations. 
This is a widely recognized challenge; a recent survey of papers using \texttt{EpiEstim} found the most common challenge for users was dealing with the quality of observed case data \citep{cori_review}. 
One sensible approach to resolving this issue is to use other sources of data. 
For instance, \citet{flaxman_2020} used a model similar to those available in \texttt{epidemia} to assess the effects of non-pharmaceutical interventions by fitting a model to death counts rather than case counts, while \citet{Mishra2020} incorporated data sources such as deaths and sero-prevalence data in addition to case data. 
Turning to other data sources is an appealing strategy for retrospective analyses, but during an ongoing epidemic it is often desirable to provide real time estimates of the effective reproduction number, a task called now-casting. 
When now-casting, case data is one of the earliest available data sources to indicate a change in the effective reproduction number. 
It behooves us, then, to develop reasonable methods for using case data when estimating the effective reproduction number, despite the difficulties involved.
\par
Our study has two main contributions. 
First, we develop our own model for estimating the effective reproduction number making different modeling choices than other available methods. 
The most significant of these is that we incorporate the number of diagnostic tests administered (both positive and negative) as a covariate in our model. 
Second, to increase understanding of the broader class of branching process inspired methods, we conduct simulation studies comparing our new model to \texttt{EpiEstim} and a model constructed using the \texttt{epidemia} package developed by \citet{epidemia}. The latter approach builds on the \texttt{EpiEstim} framework by allowing for more flexible and complex models that treat new infections as unobserved variables, with various time series such as cases or deaths modeled as noisy realizations of unobserved infections used as data \citep{epidemia, epidemia_paper}.
In particular, we explore scenarios with differing diagnostic test availability. 
We also fit our model to real data from the SARS-CoV-2 pandemic in fifteen California counties. 
Our results show that our new model outperforms existing methodologies under a variety of different testing scenarios and provides novel insights when applied to real data, highlighting the utility of incorporating tests when using case data as well as distributional choices made in the modeling process. 
\section{Methods}
\label{sec:methods}
\subsection{Available data}
Consider an outbreak observed for a total of $T$ time intervals. We restrict ourselves to two kinds of infectious disease outbreak data. The first is the time series of observed cases, $\mathbf{O} = (O_{1}, O_{2}, O_{3}, \dots, O_{T})$, where $O_{u}$ is the number of newly observed cases of an infectious disease during time interval $u$. The second is the time series of diagnostic tests, $\mathbf{M} = (M_{1}, M_{2}, \dots, M_{T})$, where $M_{u}$ is the total number of diagnostic tests administered during time interval $u$. 
For this study, we assume tests are perfectly accurate. 
We also do not model the total number of tests performed, but rather model the number of positive tests conditioned on the total number of tests. 
We assume that $O_{u}$ is a noisy realization of recent latent unobserved new infections (incidence); denoted by $I_{u}$ during time interval $u$. 

\subsection{Modeling incidence}
We first differentiate between incidence during the observation period, when case data is available, and incidence prior to the observation period. 
It is rare in practice to begin analysis of an infectious disease epidemic at the exact start of the epidemic.
We follow Scott et.\ al.\ in modeling a number of unobserved incidence values (often called seeded incidence) drawn from a hierarchical exponential model \citep{epidemia, epidemia_paper}. That is, for $t = -n, -n-1, \dots, 0$,
\begin{align*}
    \lambda &\sim \text{Exponential}(\eta), \\
    I_{t} & \sim \text{Exponential}(\lambda).
\end{align*}

We model latent incidence during the observation period as a latent gamma random variable:
\begin{align*}
I_{t}\mid \mathbf{I}_{-n:t}, R_{t} &\sim  \text{gamma}\left(R_{t}\sum_{u=-n}^{t-1} g_{t-u}I_{u}\nu, \nu\right), t=1, \dots, T,
\end{align*}
where $\mathbf{I}_{-n:t}$ is the set of all previous incidences between times $-n$ and $t$, $g_{t}$ is the discretized probability density function of the generation time (the time from an individual becoming infected to infecting someone else) distribution for the interval $t$, and $R_{t}$ is the effective reproduction number at time interval $t$.
Parameter $\nu$, describing the proportional mean-variance relationship of the above gamma distribution, receives its own prior: 
\[
\log(\nu) \sim N(\mu_{\nu}, \sigma_{\nu}^{2}). 
\]
We assume $g_{t}$ to be known. Svensson and Champredon et al. \citep{Svensson2007,champredon2015,champredon2018} have highlighted that in a closed population, the generation time distribution depends on population dynamics, i.e., it changes over time depending on the number of susceptibles available, somewhat similarly to the effective reproduction number. 
This is not taken into account in our model (nor, to our knowledge, in any model in this class of estimators). Instead we use the intrinsic generation time distribution which assumes a fully susceptible population. 
\par
Note that under this model, 
\begin{align*}
\label{eqn:mean}
    \text{E}(I_{t} \mid \mathbf{I}_{-n:t},, R_{t}) = R_{t}\sum_{u=-n}^{t-1}I_{u}g_{t-u}\tag{1},\\
    \text{Var}(I_{t} \mid \mathbf{I}_{-n:t},, R_{t}) =  R_{t}\sum_{u=-n}^{t-1}I_{u}g_{t-u}/\nu.\tag{2}
    \end{align*}
The assumed mean relationship lies at the heart of branching process inspired methods for estimating the effective reproduction number \citep{Fraser2007}. \citet{pakkanen2023unifying} show that Equation \eqref{eqn:mean} is justified under a formulation of disease transmission modeled as a variation on the Crump-Mode-Jagers branching process. 
Regardless of the underlying model, we think it is beneficial to allow for incidence to change stochastically. 
To this end, we model incidence as an auto-regressive gamma process while preserving the branching process inspired mean model \eqref{eqn:mean}. 
By modeling incidence as a continuous random variable, we are able to use Hamiltonian Monte Carlo to approximate the posterior distribution of our model parameters. 
The mean-variance relationship of the gamma distribution is also somewhat convenient, as it allows for over-dispersion in the variance of incidence through parameter $\nu$.  
\par
To allow for the effective reproduction number to change over time, we model it as a random walk on the log scale:
\begin{align*}
    \log{R_{1}} &\sim \text{Normal}(\mu_{r1}, \sigma_{r1}^{2}), \\
    \log{R_{t}} | \log{R_{t-1}}, \mathbf{I}_{-n:t-1} &\sim \text{Normal}\left(\log{R_{t-1}}, \frac{\sigma^{2}}{T-1}\right), \quad t = 2, \dots, T.\\
\end{align*}
The prior distribution of $\sigma$, $\log(\sigma) \sim N(\mu_{\sigma},\sigma_{\sigma}^{2})$, is chosen to reflect beliefs about the total amount of possible variation in the effective reproduction number over the course of the observed period. 
\par
 \subsection{Modeling observed cases}
Depending on the context of an infectious disease, the relationship between observed cases and incidence can be complex. One challenge relates to testing supply. The number of cases observed is always a function of the number of diagnostic tests administered. In the context of a novel infectious disease, testing supplies may change rapidly as new technologies are developed, approved, and deployed.  Thus, we model observed cases ($O_{t}$) conditioned on previous and current incidence ($I_{t}, \mathbf{I}_{-n:t-1},$) as a negative binomial random variable, where the mean of the negative binomial random variable is a function of incidence (as in \citep{epidemia, epidemia_paper, epinow2}), the number of tests administered, and a detection parameter $\rho$, with over-dispersion parameter $\kappa$:
\begin{align*}
    \kappa &\sim \text{Truncated-Normal}(\mu_{\kappa}, \sigma_{\kappa}^{2}), \\
    \log{\rho} &\sim \text{Normal}(\mu_{\rho}, \sigma_{\rho}), \\
     D_{t} &= \sum_{j=-n}^{t}I_{j}d_{t-j},  \\
O_{t} \mid I_{t}, \mathbf{I}_{-n:t},, \rho, \kappa, M_{t}  &\sim \text{Neg-Binom}(\rho \times M_{t} \times D_{t}, \kappa), t = 1, \dots T,\tag{2}\label{eqn:cases}
\end{align*}
where $\rho \times M_{t} \times D_{t}$ is the mean of the negative-binomial distribution. 
As defined above, $M_{t}$ is the total number of diagnostic tests administered during time interval $t$.
As a result, the detection rate for time $t$ is $\rho \times M_{t}$, which allows the detection rate to change over time as a function of the number of tests available. 
With a detection rate which depends on tests, the model can discern between situations where cases increase because of increases in latent incidence, as opposed to increases in the number of tests administered.
The weights $d_{t-j}$ are discretized weights of the delay period distribution, that is, the time from infection to detection. 
Delays occur for a variety of reasons, based on when the difference between when individuals are infected and when they test, as well as delays in reporting the results of the test.
In our study, we will use simulations and data where the only delay is caused by our assumption that cases represent individuals transitioning from the latent stage of infection to the infectious stage.
Thus, for this study, $d_{t-j}$ are discretized weights of the latent period distribution.
Note that we allow for cases observed at time $t$ to come from incidence observed at time $t$ as well; this can be adjusted depending on how quickly a particular disease spreads and at what granularity observations are recorded.
\par
It is difficult to choose generic priors for $\kappa$ and $\rho$, as they both depend in some way on properties of the surveillance system used to collect data. We address this challenge in the sections below. 

\subsection{Prior for case over-dispersion}
In our experience, some choices of the prior distribution for $\kappa$ result in poor Markov chain Monte Carlo (MCMC) convergence. To overcome this issue, we developed an approach for choosing the prior distribution for $\kappa$ inspired by Empirical Bayes methods. We fit a Bayesian thin plate regression spline to the time series of cases, assuming a negative-binomial distribution with the mean number of cases being a nonparametricaly estimated function of time, then use the posterior estimate for the over-dispersion parameter to construct the prior for our model \citep{tp_splines}. We use \texttt{brms} (version 2.15.0) to fit the regression spline to observed cases \citep{brms}. This method has drawbacks from a theoretical perspective,  because the spline-based model is fit to the same data that is then analyzed with our semi-mechanistic model. For simulations, this is easily overcome by fitting the spline to a simulated data set that is then not analyzed by our model, this is the approach we took for our simulation study. For real data analysis, one solution is to fit a spline to data from an outbreak occurring in a similar location to the one being analyzed. For this study, we put aside theoretical concerns and fit a spline-based model to each real data set used in this study to derive the prior for $\kappa$ and then applied our model. We choose the parameters of the prior by minimizing a squared loss function, searching for prior parameters which minimized the squared difference between the quantiles of the spline posterior, and the empirical quantiles of the candidate prior distribution.
\subsection{Prior for the case detection rate}
Choosing the prior for the case detection parameter $\rho$ likewise requires some care, because the meaning of $\rho$ depends on the number of diagnostic tests in the data. We propose the following procedure: first construct a plausible range for what proportion of incidence has been observed. Then, using the the 50\% quantile of tests in the observed test time series, construct a prior for $\rho$ which matches the prior for the overall mean case detection rate. In practice, we can construct the prior for $\rho$ using other quantiles as part of sensitivity analyses. For simulations we use a $\rho$ prior derived using the 50\% quantile, for real data anlysis, we use the 25\% quantile which we found improved MCMC convergence. 
\subsection{Bayesian inference}
Let $\mathbf{R} = (R_{1}, R_{2}, \dots, R_{T})$ denote the vector of effective reproduction numbers and $\mathbf{I} = (I_{-n}, \dots, I_{T})$ the vector of latent incidence counts. We are interested in the posterior distribution of our model parameters:
\begin{equation*}
    P(\mathbf{I}, \mathbf{R}, \rho, \kappa, \nu, \lambda, \sigma \mid \mathbf{O}) \propto P(\mathbf{O} \mid \mathbf{I},\rho, \kappa)P(\mathbf{I} \mid \mathbf{R}, \nu, \lambda)P(\mathbf{R} \mid \sigma)\pi(\rho, \kappa, \nu, \lambda, \sigma).
\end{equation*}
Here $P(\mathbf{O} \mid \mathbf{I},\rho, \kappa)$ defines the emissions model, $P(\mathbf{I} \mid \mathbf{R}, \nu, \lambda)$ defines the latent case model, $P(\mathbf{R} \mid \sigma)$ the random walk prior for the effective reproduction number and $\pi(\rho, \kappa, \nu, \lambda, \sigma)$ the prior on all other model parameters. 
\par
We use Hamiltonian Monte Carlo, implemented in the R package \texttt{rstan} (version 2.21.2) to approximate the above posterior distribution \citep{rstan}. For the remainder of this study we will refer to our effective reproduction number estimation method as Rt-estim-gamma.
\subsection{State-of-the-art methods}
\texttt{EpiEstim} models observed cases as incidence, and assumes that
\begin{equation*}
    I_{t} \mid I_{1}, \dots, I_{t-1}, R_{t} \sim \text{Poisson}\left(R_{t}\sum_{u=1}^{t-1}I_{u}g_{t-u}\right).
\end{equation*}
To facilitate smooth estimates, the effective reproduction number is assumed to be fixed for a given period of time, and then repeatedly estimated for all such periods in the data set. We choose a period size of one week, and allow for an uncertain generation time, re-fitting the model using different values for $g_{t-u}$ (see \citet{cori_new_2013} for details). The prior on the effective reproduction number for each window is a gamma distribution with shape parameter 1 and scale parameter 5. 
\par
Using R package \texttt{epidemia} (version 1.0.0) we created the  Rt-estim-normal model. In this model, latent incidence is an autoregressive normal random variable with variance equal to the mean multiplied by an over-dispersion parameter so the mean-variance relationship is the same as in our autoregressive gamma model. We model cases as a negative-binomial random variable, using the latent period distribution as the delay distribution (though in \texttt{epidemia} it is assumed cases cannot be generated from the current latent incidence). The case detection prior is chosen to reflect a range of plausible values for case detection depending on the simulation scenario and real data. For observed cases, we attempted to use a prior for the over-dispersion parameter that had similar values to the prior used in our model for the over-dispersion parameter of the negative binomial distribution, but found this led to issues with MCMC convergence. As such, we use the default prior for the inverse of the over-dispersion parameter implemented in \texttt{epidemia}. All other priors used are default priors from the \texttt{epidemia} package. For a full description of Rt-estim-normal, see the Appendix. 

\subsection{\texttt{EpiEstim} as an autoregressive generalized liner model}
Under the basic \texttt{EpiEstim} modeling framework, the only value in Equation \ref{eqn:mean} which is random is $R_{t}$. Consequently, \texttt{EpiEstim} can be mimicked via Poisson regression with an identity link and no intercept.
This raises the possibility of assessing the presence of over-dispersion in case data using standard statistical methods. 
To be more explicit, we can rewrite Equation \ref{eqn:mean} in the style of a generalized linear model (GLM):
\begin{equation*}
\text{E}[I_{t}\mid \mathbf{I}_{-n:t},, R_{t}] = \eta =  \beta_{1}x_{1}.
\end{equation*}
In this construction, $\beta_{1} = R_{t}$ and $x_{1} = \sum_{u=1}^{t-1}I_{u}g_{t-u}$ is the weighted sum of previous incidence. After choosing an arbitrary number of previous incidences to include in $x_{1}$, we can construct $x_{1}$ manually for every observed incidence at time $t$ with the requisite number of observed previous incidences. 
To estimate the effective reproduction number over time, we use Poisson regression repeatedly on subsets of data, where each subset has observations equal to the length of the smoothing period used in \texttt{EpiEstim}. 
For example, we can implement Poisson regression on data sets with 4 observations, estimating a $\beta_{1}$ which is fixed for those 4 observations. This is equivalent to using a period of 4 in \texttt{EpiEstim}. We assign the estimated effective reproduction number to the last date among the 4 observations, and change the settings of \texttt{EpiEstim} to match its estimates to the last observation as well. 
In addition to mimicking \texttt{EpiEstim} with a Poisson GLM, we mimic \texttt{EpiEstim} using a quasi-Poisson GLM. Using the quasi-Poisson's estimated over-dispersion parameter, we can assess how well the assumed mean variance relationship of the Poisson GLM matches the empirical variance seen in the observed data. We implement both GLM versions of \texttt{EpiEstim} and compare to the simplest version of \texttt{EpiEstim} using a fixed generation time in order to motivate the use of more complex models. All code and data needed to reproduce the results are available on GitHub at \url{https://github.com/igoldsteinh/improving_rt}.
\begin{figure}[H]
    \centering
    \includegraphics[width=\textwidth]{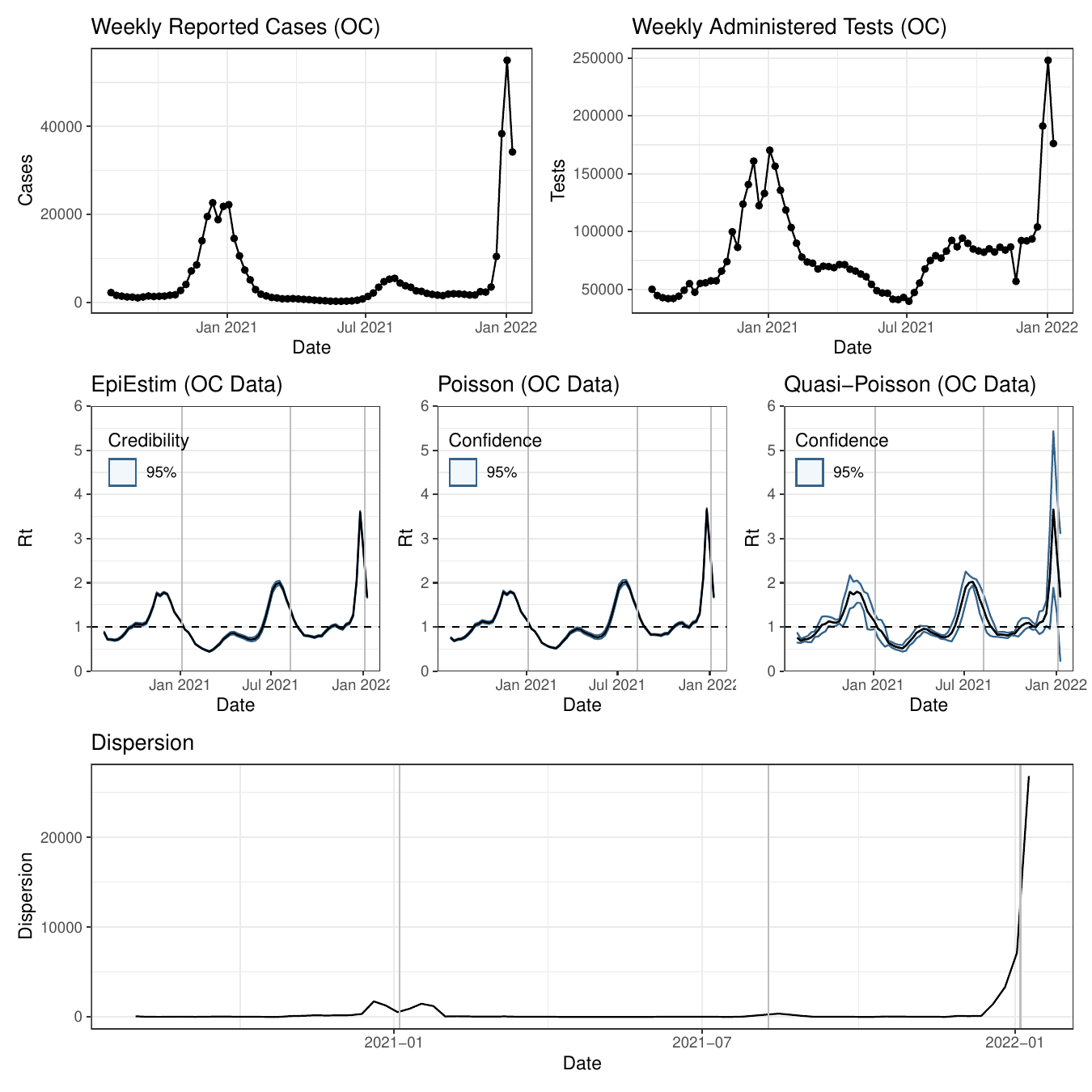}
     \caption{Estimation of the effective reproduction number of SARS-CoV-2 in Orange County, CA from Aug 2nd 2020 through January 15th 2022. The top row displays the observed cases and total diagnostic tests administered for the period. The middle row displays estimates of the effective reproduction number from \texttt{EpiEstim}, and from two GLM based mimics of \texttt{EpiEstim} using Poisson, and Quasi-Poisson regression. Blue regions represent 95\% credible or Wald confidence intervals, while black lines represent posterior median or point estimates of the effective reproduction number. The final row displays the estimated over-dispersion parameter from the Quasi-Poisson regression model. Grey vertical lines mark the date maximum statewide cases were reported for the original winter 2020 wave, the summer 2021 wave, and the winter 2021 wave.}
    \label{fig:freq_ee}
\end{figure}

\section{Results}
\subsection{GLM \texttt{EpiEstim} applied to the SARS-CoV-2 Outbreak in Orange County, CA}
To motivate the use of more complex models for estimating the effective reproduction number, we applied \texttt{EpiEstim} and our two GLM mimics of \texttt{EpiEstim} to case data from the SARS-CoV-2 outbreak in Orange County, CA from May 17th 2020 to January 15th 2021. 
We used a window size of 4 for \texttt{EpiEstim} and corresponding data sets with 4 observations for the GLM mimics. 
Data and effective reproduction number estimates are displayed in Figure \ref{fig:freq_ee}. 
The Poisson GLM closely tracks the effective reproduction number trajectory estimated by \texttt{EpiEstim}. 
However, the Quasi-Poisson estimate of the effective reproduction number has much wider confidence intervals than the Poisson GLM. This is because the estimated over-dispersion parameter in the Quasi-Poisson model ranges from 1.01 to 26851.84. 
This shows that the Poisson model for incidences may be inadequate, resulting in overconfidence of $R_t$ inference.

\subsection{Simulation protocol}
Simulated data for this study was generated from a stochastic SEIR model in R (version 4.0.4) using the \texttt{stemr} package (version 0.2.0) \citep{R,stemr}. SEIR models generate an infectious disease outbreak at a population level, with the population divided into four compartments: susceptible, exposed (infected but not yet infectious), infectious, and removed (neither infectious nor susceptible). The changes in these compartments are governed by rate parameters which depend on the populations in the compartments. In our simulations, the mean latent period was 4 days, the mean infectious period was 7.5 days. Daily case data was generated from transitions from the E to the I compartment on day t, using a fixed number of tests and a negative binomial distribution. For all simulations, $\rho$ was set to be $9\times 10^{-5}$ and $\kappa$ was set to be 5.
\par
The basic reproduction number $R_{0}$ was given a fixed trajectory, leading to similar $R_{t}$ trajectories for each realization of the simulation. More details on the stochastic SEIR model used for simulation are available in Appendix section \ref{app:sim}. 
Note that the SEIR models used for the simulations do not match any of the models used for inference of the $R_t$ trajectories. 
In other words, all our simulation results are produced in the presence  of model misspecification ---  a desirable feature for a realistic simulation protocol. 
\par
\begin{figure}[H]
    \centering
    \includegraphics[width = \textwidth]{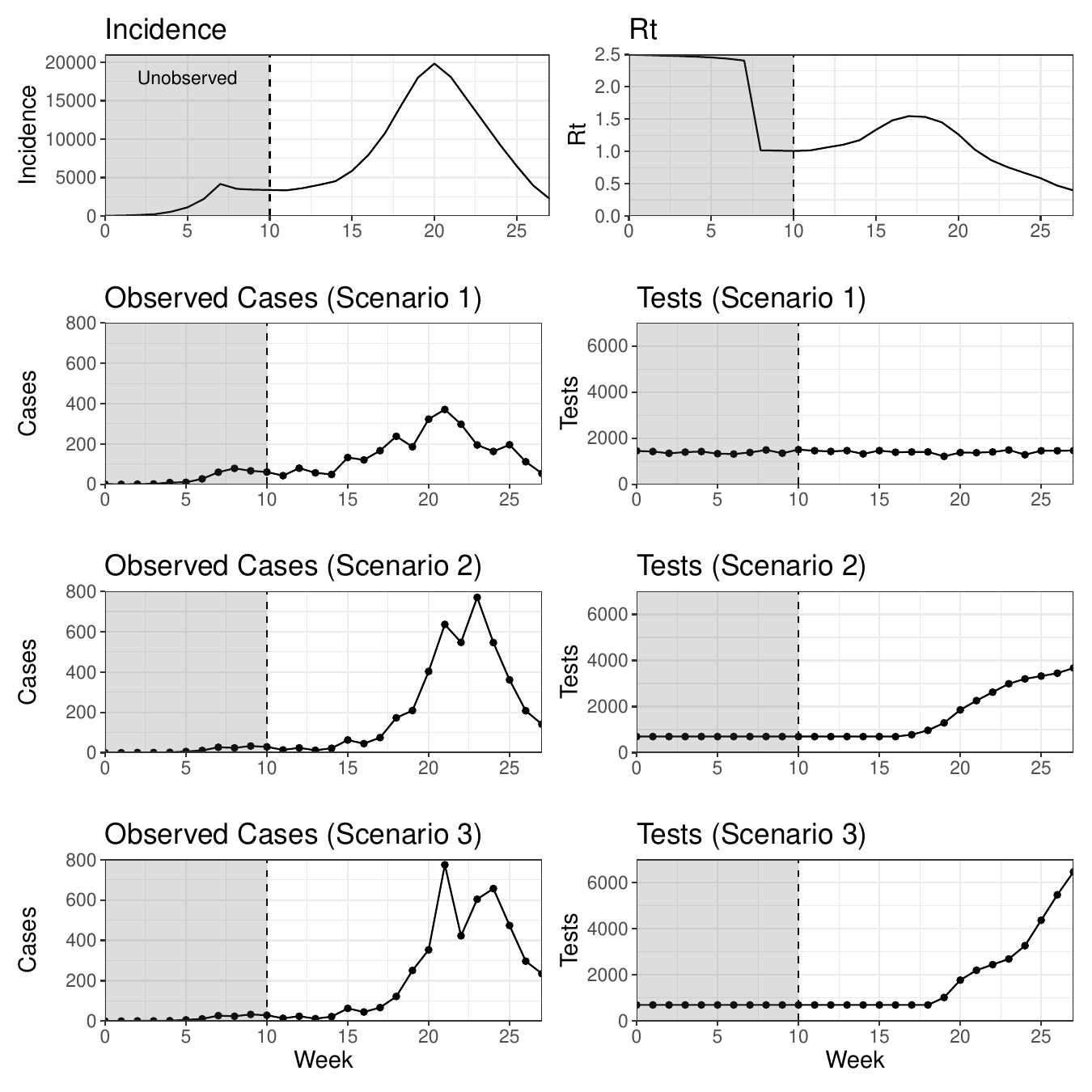}
    \caption{Simulated epidemic data generated from an SEIR model. Cases are generated using an emissions model which includes total diagnostic tests administered as a covariate. Three different testing scenarios are considered, all with the same underlying R0 trajectory. Included are underlying incidence and effective reproduction number trajectories. While these may vary slightly across simulations, they will be very similar due to identical infectious disease dynamics. In Scenario 1, tests are randomly sampled from a normal distribution. In Scenarios 2 and 3, tests stay flat and then increase at varying rates. Simulated epidemics start with 10 individuals and last for 28 weeks. The first 11 weeks are discarded, and are not used when simulated data are analyzed by the three effective reproduction number estimation methods.}
    \label{fig:simfig}
\end{figure}

We simulated three separate scenarios lasting 28 weeks, where all parameters were the same except for the number of tests at each time step. 
In Scenario 1, weekly tests were drawn from a normal distribution with parameters that remained constant over time. 
In Scenario 2, tests were held constant for the first six weeks of the simulation, then increased at varying rates over the next eleven weeks of the simulation. Scenario 3 was similar to Scenario 2, except that testing was held constant for the first eight weeks, and increased more quickly than in Scenario 2. All simulations were done on a daily time scale, then aggregated into weeks for analysis. 
The true effective reproduction number for a single week was taken to be the true effective reproduction number of the third day of that week. 
In all simulations, the first 11 weeks were not analyzed, leaving 17 weeks of data for analysis. 
For each scenario, we generated 100 simulations.
Realizations of all three simulations are displayed in Figure \ref{fig:simfig}. 

\begin{table}
\caption{Priors used by the Rt-estim-gamma method in the  simulation study.}
\centering
\fbox{%
\begin{tabular}{*{4}{c}}
          Parameter & Simulation & Prior & Prior Median (95\% Interval)  \\
         \hline \\
         $\nu$ & All &  Log-normal(-2, 0.7) & 0.15 (0.03, 0.53) \\
         $\sigma$ & All &  Log-normal(-0.66, 0.6) & 0.52 (0.16, 1.68) \\
         $\lambda$ & All & Exponential(0.3) &  2.31 (0.08, 12.26) \\
         $\log{R_{1}}$ & All &  Normal(0, 0.75) & 0.01 (-1.49, 1.49) \\ 
         $\rho$ & Scenario 1 & Log-normal(-11.06, 0.3) & 1.57E-5 (8.756E-6, 2.85E-5) \\
         $\rho$ & Scenario 2 &  Log-normal(-11.43, 0.3) & 1.09E-5 (5.96E-6, 1.96E-5) \\
         $\rho$ & Scenario 3 &  Log-normal(-11.56, 0.3) & 1.57E-5 (8.81E-6, 2.83E-5) \\
         $\kappa$ & Scenario 1 &  Truncated-Normal(59, 60) & 72.00 (5.00, 183.15) \\
         $\kappa$ & Scenario 2 &  Truncated-Normal(33, 25) & 35.65 (3.14, 83.23) \\
         $\kappa$ & Scenario 3 & Truncated-Normal(70, 80) & 88.84 (6.00, 235.41)
\end{tabular}}
\label{tab:sim_priors} 
\end{table}
\newpage
\subsection{Simulation results}
\begin{figure}[H]
    \centering
    \includegraphics[width = \textwidth]{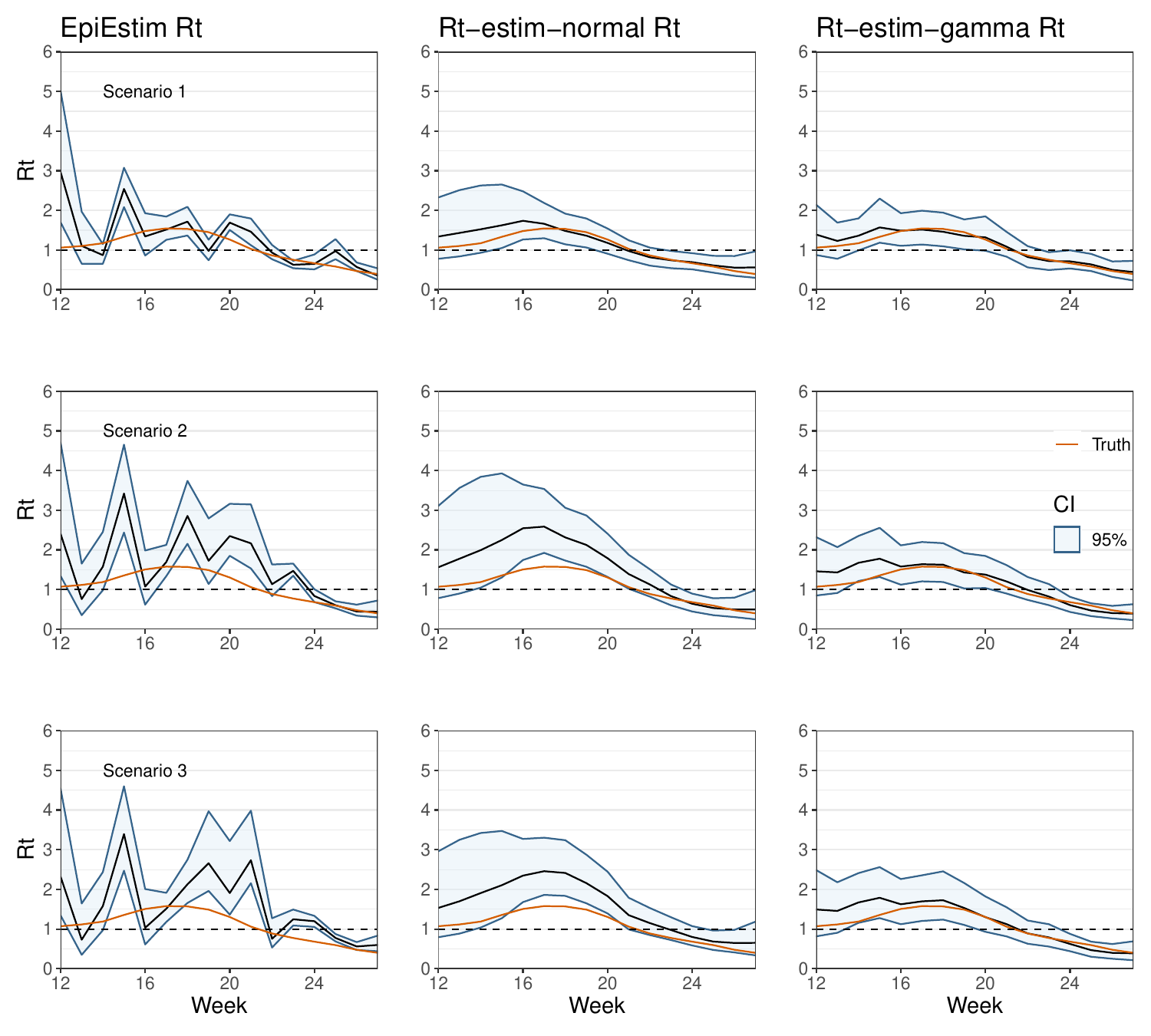}
    \caption{$R_{t}$ estimation using three different methods for three simulated data sets under different testing scenarios. True $R_{t}$ trajectories are colored in red, black lines represent median estimates from the posterior distribution, blue shaded areas are 95\% credible intervals.}
    \label{fig:sim_rt}
\end{figure}

For each model fit using \texttt{rstan}, we sampled 2000 posterior draws, discarding the first half as burn-in. Figure \ref{fig:sim_rt} visualizes the estimates for the effective reproduction number from \texttt{EpiEstim}, Rt-estim-normal and Rt-estim-gamma for the three data sets visualized in Figure \ref{fig:simfig}. 
We checked convergence diagnostics for Rt-estim-normal and Rt-estim-gamma for all simulations and ensured adequate convergence of all models. More details are in the Appendix section \ref{convergence}.
Since \texttt{EpiEstim} does not provide estimates for the first time point in the series, we report only time points for which all three methods have estimates. 
\par
Credible intervals for \texttt{EpiEstim} frequently miss the true $R_{t}$ values (covering between 6 and 9 of the 16 true values), while credible intervals for Rt-estim-normal and Rt-estim-gamma cover most true values across simulations. However, Rt-estim-gamma covers more true values than Rt-estim-normal, with narrower credible intervals (ranging between 11 and 16 values for Rt-estim-normal, and 16 values for every scenario for Rt-estim-gamma). 
\par
\begin{figure}[H]
    \centering
    \includegraphics[width = \textwidth]{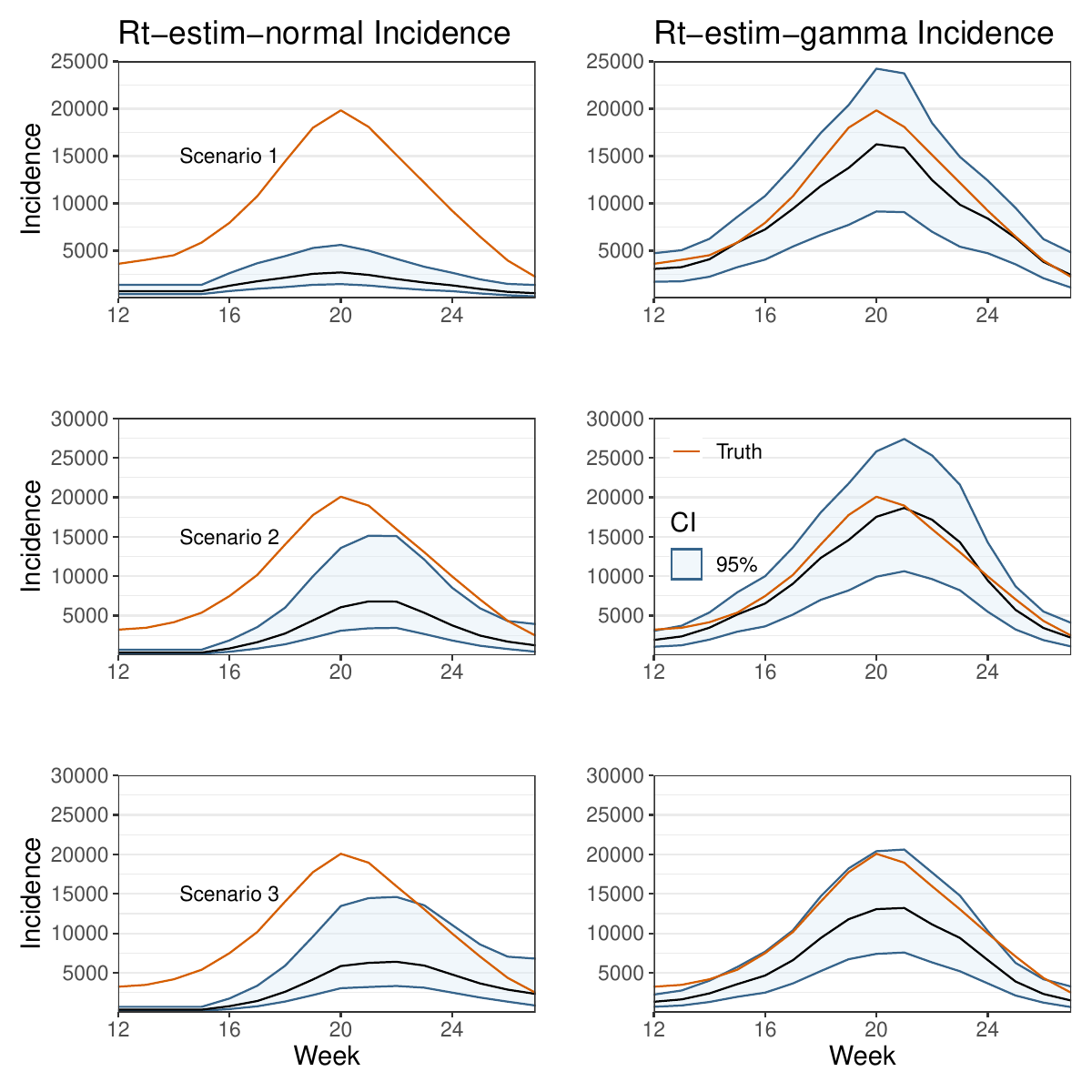}
    \caption{Incidence estimation using two different $R_{t}$ estimation methods for three simulated data sets under different testing scenarios. True incidence trajectories are colored in red, black lines represent median estimates from the posterior distribution, blue shaded ares are 95\% credible intervals.}
    \label{fig:sim_incid}
\end{figure}

Figure \ref{fig:sim_incid} visualizes estimates of latent incidence from Rt-estim-normal and Rt-estim-gamma for the three data sets visualized in Figure \ref{fig:simfig}. 
Rt-estim-normal credible intervals rarely cover the true incidence (covering from 0 to 5 to true values), while Rt-estim-gamma credible intervals generally do (covering 11 to 16 true values).
\par
Posterior predictive distributions for cases for both Rt-estim-normal and Rt-estim-gamma  are displayed in Appendix Figure \ref{fig:sim_case} (the posterior predictive distribution for \texttt{EpiEstim} is not readily available). For all three scenarios, for both models, 95\% credible intervals from the posterior predictive distributions cover all observed data points. Rt-estim-gamma had generally narrower credible intervals than Rt-estim-normal. 
\begin{figure}[H]
    \centering    
    \includegraphics[width = \textwidth]{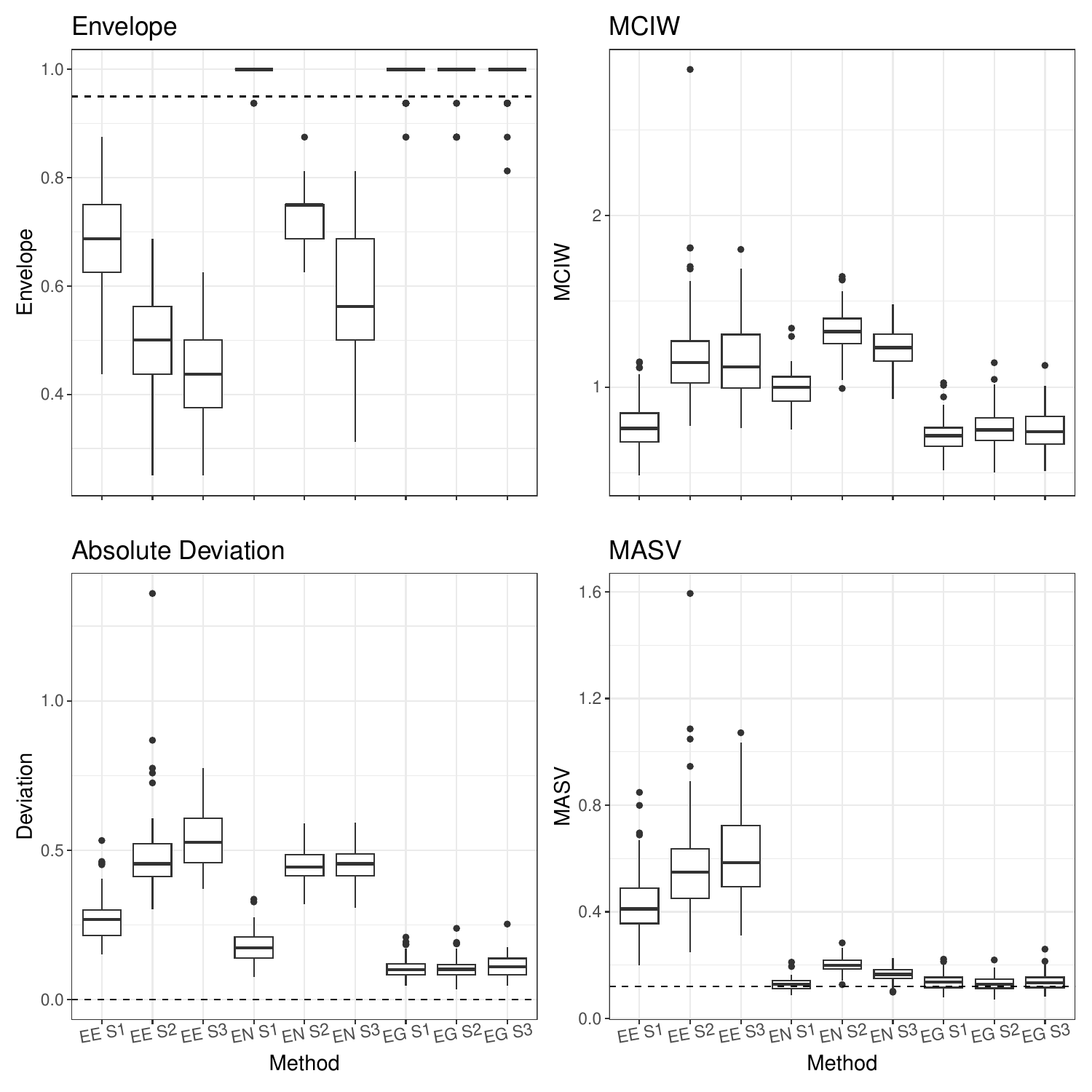}
    \caption{Frequentist metrics for \texttt{EpiEstim}, Rt-estim-normal, and Rt-estim-gamma applied to three different simulated epidemics. Envelope is a measure of coverage, taking the average coverage of 95\% intervals over the time series. MCIW is the average mean credible interval width. Absolute deviation is the mean of the absolute value of the difference between the median $R_{t}$ at each time point and the true value. Mean absolute standard deviation is the difference between the current median point estimate for $R_{t}$ and the previous point estimate for $R_{t}$. The dashed line in the bottom right panel represents the true absolute standard deviation. Solid lines represent medians, hinges are upper and lower quartiles and whiskers are at most 1.5 times the inter-quartile range from the median. \texttt{EpiEstim} is denoted EE, Rt-estim-normal EN, and Rt-estim-gamma EG, while scenarios are marked S1, S2 and S3. EE S1 describes results of using \texttt{EpiEstim} in simulation scenario 1.}    \label{fig:freq_metric}
\end{figure}
Because we are using a stochastic SEIR model to generate simulations, each simulation has a different, though similar in shape, true effective reproduction number curve (despite having the same true basic reproduction number curve).
The range of true effective reproduction number curves is visualized in Figure \ref{fig:s1_truert}.
We report frequentist metrics in order to summarise performance across a variety of different epidemic curves.
Model performance on simulated data sets for each of the three models is summarized in Figure \ref{fig:freq_metric}. For each metric, we summarize results in boxplots where solid lines represent medians, hinges are upper and lower quartiles and whiskers are at most 1.5 times the inter-quartile range from the median. Envelope is a measure of coverage. For each simulation the envelope is the proportion of time points for which a 95\% credible interval from the posterior distribution captured the true value of interest. Mean credible interval width (MCIW) is the mean of credible interval widths across time points within a simulation. Absolute deviation is a measure of bias, and is the mean of the absolute difference between the posterior median and the true value at each time point. Finally, mean absolute sequential variation (MASV) measures how well each method captured the variation in the effective reproduction number across time by computing the mean of the absolute difference between the posterior median at time point $t$ and the posterior median at time point $t-1$. We compare this to the true mean absolute sequential variation in each simulation. \texttt{EpiEstim} had the lowest envelope in all simulation scenarios. Rt-estim-normal had high envelope in Scenario 1 but dropped to lower values in scenarios with time-varying testing supply. Rt-estim-normal had the largest MCIW in all three scenarios, while Rt-estim-gamma had the smallest MCIW in all three scenarios. \texttt{EpiEstim} and Rt-estim-normal had relatively similar values for absolute devaition, Rt-estim-gamma had the smallest absolute deviation in all scenarios. Finally, \texttt{EpiEstim} had the largest MASV in all scenarios, while Rt-estim-normal and Rt-estim-gamma had relatively comparable MASV. For two of three scenarios, Rt-estim-gamma was closer to the true MASV than Rt-estim-normal. 
We ran three additional experiments using the data sets from Scenario 3 to better understand our model. All results are displayed in Appendix Figure \ref{fig:app_freq_metric}, with the results from Figure \ref{fig:freq_metric} included as a baseline comparison. In the first experiment, we halved each parameter in the hypo-exponential distribution and refit the model to the data sets from Scenario 3. This led to narrower credible intervals and lower envelope (see Appendix Figure \ref{fig:app_freq_metric} for results). In the second experiment, we used a spline fit to the same data being analyzed in order to choose a prior for $\kappa$. We found no meaningful difference in performance. In the third experiment, we used a prior for $\rho$ derived from the 25\% quantile of tests, rather than the 50\% quantile, this again led to no meaningful difference in performance with regards to estimating the effective reproduction number, though we expect it to change estimates of incidence. 
Overall, we find that our Rt-estim-gamma model outperforms the \texttt{EpiEstim} and Rt-estim-normal in all metrics, surprisingly  even in Scenario 1, where the number of diagnostics tests did not vary appreciably over time (Rt-estim-gamma  and Rt-estim-normal have similarly high envelope values in this case).

\subsection{Estimating the effective reproduction number of SARS-CoV-2 in fifteen California counties}
We analyzed SARS-CoV-2 reported case data from fifteen California counties representing Northern California (Alameda, Sacramento, San Francisco, Santa Clara, Contra Costa), Central California (Fresno, Merced, Monterey, Stanislaus, Tulare) and Southern California (Los Angeles, Orange, Riverside, San Bernardino, San Diego). These counties represent more than 75 percent of the population of California, and differ widely along demographic, economic, and political characteristics. We analyzed data from August 2nd 2020 through January 15th 2022. Data are publicly available from the California Open Data Portal \citep{caopendata}. Positive cases are associated with the date of their test, rather than the date they were reported.
\par
To estimate the effective reproduction number of SARS-CoV-2 we must choose a generation time distribution to use in our models. Estimating intrinsic generation times from observed data is non-trivial \cite{Park2021}. Early efforts from \citet{Ferretti2020} and \citet{Ganyani2020} estimate the mean generation time to be between 5.5 and 5.2 days respectively. A more recent estimate of the mean intrinsic generation interval for the original version of SARS-CoV-2 estimated it to be 9.7 days \citep{Sender2021}, but the issue of optimal generation time inference seems far from settled. An additional complication is that a number of important variants of SARS-CoV-2 have spread over the course of the pandemic, and the generation times for the variants may differ from that of the original viral strain. 
\citet{hart2022} found it is likely that the intrinsic mean generation time of the delta variant is shorter than that of the alpha variant, likewise a preliminary study by Abbott et al. suggests the intrinsic mean generation time of the omicron variant is shorter than that of the delta variant \citep{abbott2022}. 
We find the methodology of Sender et al. somewhat persuasive, and use their point estimate of the generation time (a log-normal distribution with  mean 9.7 days) as the default generation time for the original SARS-CoV-2 strain and alpha variant versions of SARS-CoV-2. 
We compare these default findings to results using the Ferretti et al. point estimate distribution (a Weibull distribution with mean 5.5 days) in Appendix A.6. 
\par
We then created an alternative version of our model which allowed for changing the generation time distribution due to the delta and omicron variants. We changed the generation time distribution starting in July 2021, reflecting our assumption that delta variant dominated new cases by this point, and changed it again in December 2021, reflecting the same assumption about the omicron variant. Hart estimates the median reduction in the mean generation time for the delta variant is 15\% as compared to alpha (we assumed alpha and wild-type had the same generation time) \citep{hart2022}, while Abbott estimates the median reduction in the mean generation time for omicron is 28\% as compared to delta \citep{abbott2022}. We created generation time distributions for these variants by minimizing a squared loss function to search for parameters such that the new distributions had the appropriate new mean generation time, while preserving the standard deviation of the original distribution (see Appendix for complete details). 
We tested whether this new model was needed by calculating the Bayes factor of the two models using data from Alameda County the \texttt{bridgesampling} package in R \citep{meng1996simulating, bridgesampling}, running both models for 26,000 iterations with the first 1000 iterations discarded as burn in on 3 chains. The point estimates for the marginal likelihood had error of 7\% for the constant generation time model and 6\% for the varying generation time model, with a reported Bayes Factor of 1.58 in favor of the model with variant-specific  generation times. 
Even accounting for the margin of error, it is hard to conclude the varying generation time model was decisively superior to the constant generation time model, so we used the constant generation time model in this paper. Because we were testing a characteristic of the infectious disease which should generalize across locations, and because of the computational cost involved, we did not calculate Bayes factors for all fifteen counties.
\par
Finally, we used the point estimate of the latent period distribution from \citet{Xin2021} as the delay distribution in our model, with a mean latent period of 5.5 days using a gamma distribution. For the alternative analysis using the Ferretti et al. distribution, we scaled this distribution by 0.5 to halve the mean latent period. 
\par
\begin{figure}[H]
    \centering
    \includegraphics[width = \textwidth]{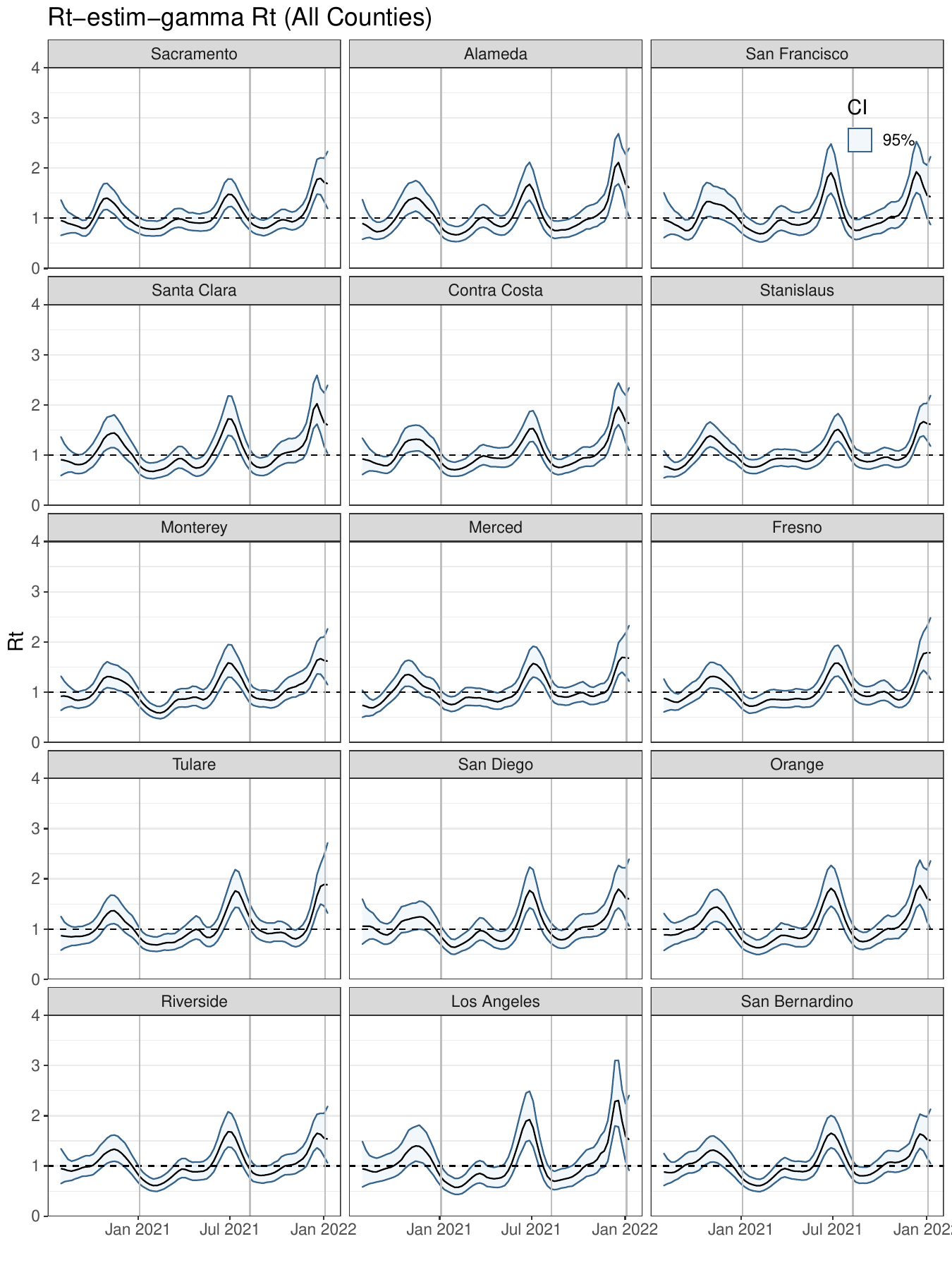}
    \caption{Estimates of the effective reproduction number of SARS-CoV-2 from Rt-estim-gamma applied to fifteen counties in California, USA from August 2nd 2020 through January 15th 2022. Blue shaded regions are 95\% posterior credible intervals. Black lines are medians. Grey vertical lines mark the date maximum statewide cases were reported for the original winter 2020 wave, the summer 2021 wave, and the winter 2021 wave.}
    \label{fig:allcounties_rt}
\end{figure}

 
We fit \texttt{EpiEstim}, Rt-estim-normal (using the priors from the simulations), and Rt-estim-gamma (see Appendix for priors) to this data. 
The posterior summaries for the effective reproduction number as calculated by \texttt{EpiEstim} are displayed in Figure \ref{fig:allcounties_epiestim} and those calculated by Rt-estim-gamma are displayed in Figure \ref{fig:allcounties_rt}. 
Accompanying incidence posterior distributions and case posterior predictive distributions for Rt-estim-gamma are displayed in Figures \ref{fig:allcounties_incid} and \ref{fig:allcounties_cases} respectively. 
Visualizations of the priors and posteriors for non time-varying parameters for Rt-estim-gamma fit to Los Angeles County data are displayed in Figure \ref{fig:la_prior_post}. 
After running into convergence issues with Rt-estim-normal, we reduced the data set to August 2nd 2020 through November 6th 2021 and fit Rt-estim-normal to this data set. The Rt-estim-normal results were generated using R version 4.2.2.
\par
Priors for Rt-estim-gamma were the same as in the simulations, except that the prior $\sigma$ had a mean of -0.61 (the range of plausible values was similar), and the priors for $\rho$ and $\kappa$ were chosen for each county individually using the protocols described in the methods section. We assumed the overall median proportion of observed incidence was 0.066. An example of posterior predictive intervals from the thin plate spline used to choose the prior for $\kappa$ and from Rt-estim-gamma fit to SARS-CoV-2 case data from Alameda County, California, are visualized in Figure \ref{fig:spline_example}.
\par 
Comparisons with Rt-estim-normal are displayed in Figures \ref{fig:ca_rt_estimnormal}, \ref{fig:ca_incid_estimnormal} and \ref{fig:ca_cases_estimnormal}. The Rt-estim-gamma results were generated in R version 4.1.2, but all packages were the same as those used to generate simulation results except for \texttt{Rcpp} which was version 1.0.8 rather than version 1.0.7. 
Overall, Rt-estim-gamma estimates were smoother and more uncertain than estimates from \texttt{EpiEstim}, but less smooth and uncertain than those produced by Rt-estim-normal. This behavior is consistent with model performance in the simulation scenarios. Rt-estim-gamma estimates tended to estimate less extreme magnitudes than Rt-estim-normal estimates, and while the two models produced broadly similar estimates of the trajectory of the effective reproduction number, they differed in some counties in significant ways. For example, in San Diego county, the median estimate from Rt-estim-normal is always above 1 before January 2021, while the median estimate from Rt-estim-gamma is below 1 for parts of this period. Additionally, in all counties the median estimate from Rt-estim-gamma crossed below 1 before the median estimate for Rt-estim-normal in fall 2021. Rt-estim-normal and Rt-estim-gamma produced different estimates of the latent incidence (Figure \ref{fig:ca_incid_estimnormal}), but both produced 95\% posterior predictive intervals for the observed cases which had good coverage in all counties (Figure \ref{fig:ca_cases_estimnormal}). 
\par
The results using Rt-estim-gamma with a mean generation time of 5.5 days are displayed in Figures \ref{fig:allcounties_rt_shortgen}, \ref{fig:allcounties_incid_shortgen} and \ref{fig:allcounties_cases_shortgen}. Using a shorter generation time led to generally smaller estimates of the peak effective reproduction number with narrower credible intervals. However, the trajectories using either generation time were similar, and the estimated trajectories agreed on when the median reproduction number was above or below one. 
\par
Median estimates for the effective reproduction number were larger during the summer 2021 wave than during the winter 2020 wave. The estimate of the reproduction number during the winter 2021 wave was similar to that of summer 2021 wave except in a few counties where it was larger, such as Los Angeles and Alameda counties. Trajectories were similar across counties, but varied in timing and magnitude from county to county. For instance, the peak reproduction number in the Winter 2020 wave was estimated to occur in the week of November 1st in Sacramento County, and the week of November 22nd in Los Angeles County.  

\section{Discussion}
We presented a model for estimating the effective reproduction number using time series of observed cases and diagnostic tests, as well as methods for choosing key priors for the model. We tested the model on simulated data sets, showing it can successfully estimate the true effective reproduction number when data is generated from a stochastic compartmental model. We also tested other models used for estimating the effective reproduction number, demonstrating that when testing supply is relatively constant, a case observation model which ignores testing is reasonable, but when testing supply changes rapidly, ignoring testing leads to poor model performance. Using data from the SARS-CoV-2 epidemic in California, we have shown how using a model fit to case observation data that incorporates testing data leads to different conclusions about the trajectory and magnitude of the effective reproduction number in real world epidemics.
\par
We found that \texttt{EpiEstim} had poor performance across all simulation scenarios. In contrast, an assessment of \texttt{EpiEstim} by Gostic et.\ al found it had reasonable performance on simulated data and recommended it over other existing methodologies (models available in \texttt{epidemia} were not assessed in this study) \citep{gostic2020}. However, Gostic et.\ al.\  only tested \texttt{EpiEstim} on simulated data sets where the true incidence was known. In our study, we tested \texttt{EpiEstim} on data sets where cases were noisy realizations of unobserved incidence, a much more realistic scenario for many diseases, such as SARS-CoV-2. The performance of our GLM versions of \texttt{EpiEstim} on data from the SARS-CoV-2 epidemic in Orange County, CA provides one reason for this poor performance. Modeling reported cases as a Poisson random variable assumes a stringent mean-variance relationship which is likely to under-estimate uncertainty. We do not recommend using \texttt{EpiEstim} to estimate the effective reproduction number when there is reason to believe reported cases do not reflect true incidence.
\par 
When testing was relatively constant, the Rt-estim-normal model, which assumes latent incidence but ignores tests, still performed well. Even in this scenario, the Rt-estim-gamma model we developed for this study had smaller mean credible interval widths and smaller absolute deviations. This suggests that our other modeling choices beyond including tests as a covariate, such as the use of the gamma distribution to model latent incidence, and our process for choosing the prior for the case over-dispersion parameter, had positive effects on model performance. As the SARS-CoV-2 pandemic has unfolded, a number of modeling groups have developed similar techniques for estimating the effective reproduction number. 
We have demonstrated how modifying distributional assumptions and developing protocols for choosing priors can have significant impact on model performance. We hope these findings motivate the larger community of researchers focused on modeling the effective reproduction number to revisit their work and establish best practices for this class of models. 
\par
In simulation scenarios where testing supply increased dramatically, we were still able to successfully estimate the effective reproduction number using the Rt-estim-gamma model. Our findings suggest incorporating testing data is a viable strategy for using case data, which should improve the accuracy of efforts at now-casting the effective reproduction number. It is worth noting that we avoided using a delay distribution in Rt-estim-gamma which incorporated reporting delays, instead using data where cases and tests were tied to the date of the test. 
This should not prevent the use of Rt-estim-gamma for up-to-date now-casting even though counts of the most recent cases and tests will inevitably be under-counts, so long as the proportion of positive to total tests is independent of reporting delays. We assumed this was the case when applying Rt-estim-gamma to the SARS-CoV-2 data from California. 
If this assumption proves not to be true, then approaches which do not use testing data and incorporate more elaborate delay distributions, such as those of \citet{abbott_estimating_2020} and \citet{epidemia_paper} are probably a better choice.
Similarly, because our model relies on the proportion of positive to total tests, rather than the raw counts of positive tests, it should be robust to changes in types of tests available, so long as reported positive cases used the same kinds of tests recorded in total diagnostic tests. This allows us to avoid any problems which arise from the availability of rapid tests for SARS-CoV-2 during the omicron wave.
In a representative set of simulations, even when Rt-estim-normal posteriors captured the effective reproduction number, its posterior estimates for latent incidence did not capture the true latent incidence. To a lesser extent, the same was true of the Rt-estim-gamma model. We have not yet seen any discussion as to the accuracy of incidence estimation for this class of models. Our findings suggest incidence estimates should not be trusted, as there are many values for incidence which lead to the same observed cases and the same reproduction number estimates. Estimates where a population size are taken into account, such as in \citet{Mishra2020}, may be more trustworthy, but we recommend running a simulation study first to verify this.
\par
One important limitation of our method is that we condition on the number of tests and use them as a covariate, rather than modeling them jointly with cases.
We would expect in practice that the number of tests is also a function of past incidence (with cases rising, more tests will be administered). 
In mathematical terms, a joint model of cases and tests could be written as 
\begin{equation*}
        P(\mathbf{O}, \mathbf{M} \mid \mathbf{I}, \rho, \kappa) = P(\mathbf{O} \mid \mathbf{M}, \mathbf{I}, \rho, \kappa)P(\mathbf{M} \mid \mathbf{I}, \rho, \kappa).
\end{equation*}
Our method only uses the first term of this product. 
This leaves our method open to potential bias from model misspecification.
In the simpler context of regression without latent variables, this issue is sometimes called ``feedback", a thorough treatment of the topic is available in Chapter 12 of \citet{diggle2002analysis}.
While we think that in practice this will not be a concern in situations where cases and tests increase and decrease together in response to changes in incidence, the possibility does exist. 
For instance, suppose the testing policy during the peak of an epidemic was that individuals with symptoms could not be tested, as anyone with symptoms should simply assume they have been infected. 
Tests might still increase in response to increased incidence from the wave, but cases could decline, because no symptomatic people were testing. 
In such a scenario, we would expect our model to fail. 
Modeling tests  is a non-trivial problem, and implementing a joint model of cases and tests is a promising future direction.

\par
In this paper, we used gamma densities to in order to model changes in latent incidence stochastically. 
While our choice of a gamma distribution has some desirable benefits, namely that it allows us to use HMC to generate posterior samples, and that it allows for overdispersion in the variance, there is definitely room for improvement in modeling latent incidence.
Recent work by \citet{penn2022uncertainty} provides an interesting avenue for improvement, with explicit calculations of the variance of the transition distributions of a time-varying general branching process.
\par 
In the real data analysis, we used Rt-estim-gamma with a prior for the over-dispersion parameter derived from a spline fit to the same data as Rt-estim-gamma. 
This is a workaround we developed to avoid computational problems related to using Hamiltonian Monte Carlo when the prior for the over-dispersion parameter strongly conflicts with the data. 
Another MCMC method, such as Zig-Zag sampling \citep{bierkens2017, corbella2022automatic}, may not have this issue, and so we could avoid this procedure.
While not ideal, we tested our model using this procedure for choosing the over-dispersion prior on simulated data, and found no discernible loss in performance.
\par
It is clear that more sophisticated representations of the generation time distribution which could change according population dynamics could be incorporated into our model. While this might lead to improved model performance, it is encouraging that in simulations, our model performed well despite using a fixed generation time. It is equally encouraging that our experiments on both simulated and real data showed our model was reasonably robust to different generation time distributions.   
\par
One obvious area for improvement in this space is allowing the prior on case detection ($\rho$) to change over time to better reflect changes in testing policy. 
For instance, at the start of the SARS-CoV-2 pandemic, only symptomatic individuals could get tested in California, whereas in Fall 2021, anyone was eligible to receive a test. 
\citet{sherratt2021exploring} also highlighted how changes in testing eligibility may result in estimating spurious changes in the effective reproduction number. We have found case data alone is insufficient to make a time varying detection parameter identifiable. Incorporating other sources of data which facilitate real-time estimation, such as data from wastewater treatment facilities, may enable models with time varying case detection parameters. Enabling effective reproduction number estimation methods to incorporate multiple data streams seems like a fruitful area of future research. 

\section*{Acknowledgments}
The authors are grateful to Jon Fintzi and Damon Bayer for their help using the \texttt{stemr} package. We are grateful for funding from the UCI Infectious Disease Science Initiative.  
This work utilized the infrastructure for high-performance and high-throughput computing, research data storage and analysis, and scientific software tool integration built, operated, and updated by the Research Cyberinfrastructure Center (RCIC) at the University of California, Irvine.

\section*{Funding}
This work was made possible in part through support from the UC CDPH Modeling Consortium. I.H.G and V.N.M were supported in part by NIH grant R01AI147336. J.W was supported in part by NIH grant NIH R01AI029168.

\section*{Data Availability}
 All data needed to reproduce the results are available on GitHub at \url{https://github.com/igoldsteinh/improving_rt}.
\newpage
\bibliographystyle{../../jrssa-submission/rss.bst}  
\bibliography{../../references}  

\begin{thebibliography}{42}
\expandafter\ifx\csname natexlab\endcsname\relax\def\natexlab#1{#1}\fi
\expandafter\ifx\csname url\endcsname\relax
  \def\url#1{\texttt{#1}}\fi
\expandafter\ifx\csname urlprefix\endcsname\relax\def\urlprefix{URL: }\fi

\bibitem[{Abbott et~al.(2020{\natexlab{a}})Abbott, Hellewell, Thompson, Sherratt, Gibbs, Bosse, Munday, Meakin, Doughty, Chun, Chan, Finger, Campbell, Endo, Pearson, Gimma, Russell, {CMMID COVID modelling group}, Flasche, Kucharski, Eggo and Funk}]{abbott_estimating_2020}
Abbott, S., Hellewell, J., Thompson, R.~N., Sherratt, K., Gibbs, H.~P., Bosse, N.~I., Munday, J.~D., Meakin, S., Doughty, E.~L., Chun, J.~Y., Chan, Y.-W.~D., Finger, F., Campbell, P., Endo, A., Pearson, C. A.~B., Gimma, A., Russell, T., {CMMID COVID modelling group}, Flasche, S., Kucharski, A.~J., Eggo, R.~M. and Funk, S. (2020{\natexlab{a}}) Estimating the {Time-Varying} {Reproduction} {Number} of {SARS}-{CoV}-2 {Using} {National} and {Subnational} {Case} {Counts}.
\newblock \textit{Wellcome Open Research}, \textbf{5}, 112.
\newblock \urlprefix\url{https://wellcomeopenresearch.org/articles/5-112/v2}.

\bibitem[{Abbott et~al.(2020{\natexlab{b}})Abbott, Hellewell, Thompson, Sherratt, Gibbs, Bosse, Munday, Meakin, Doughty, Chun et~al.}]{epinow2}
Abbott, S., Hellewell, J., Thompson, R.~N., Sherratt, K., Gibbs, H.~P., Bosse, N.~I., Munday, J.~D., Meakin, S., Doughty, E.~L., Chun, J.~Y. et~al. (2020{\natexlab{b}}) Estimating the time-varying reproduction number of {SARS-CoV-2} using national and subnational case counts [version 2; peer review: 1 approved with reservations].
\newblock \textit{Wellcome Open Res 2020, 5:112}, \textbf{5}, 112.
\newblock \urlprefix\url{https://doi.org/10.12688/wellcomeopenres.16006.2}.

\bibitem[{Abbott et~al.(2022)Abbott, Sherratt, Gerstung and Funk}]{abbott2022}
Abbott, S., Sherratt, K., Gerstung, M. and Funk, S. (2022) Estimation of the test to test distribution as a proxy for generation interval distribution for the {Omicron} variant in {England}.
\newblock \textit{medRxiv}, 2022.01.08.22268920.

\bibitem[{Bhatt et~al.(2023)Bhatt, Ferguson, Flaxman, Gandy, Mishra and Scott}]{epidemia_paper}
Bhatt, S., Ferguson, N., Flaxman, S., Gandy, A., Mishra, S. and Scott, J.~A. (2023) {Semi-Mechanistic {B}ayesian modeling of {COVID}-19 with Renewal Processes}.
\newblock \textit{Journal of the Royal Statistical Society Series A: Statistics in Society}, in press.

\bibitem[{Bierkens and Roberts(2017)}]{bierkens2017}
Bierkens, J. and Roberts, G. (2017) A piecewise deterministic scaling limit of lifted {M}etropolis-{H}astings in the {C}urie-{W}eiss model.
\newblock \textit{The Annals of Applied Probability}, \textbf{27}, 846 -- 882.

\bibitem[{Bürkner(2017)}]{brms}
Bürkner, P.-C. (2017) {brms}: An {R} {Package} for {Bayesian} {Multilevel} {Models} {Using} {Stan}.
\newblock \textit{Journal of Statistical Software}, \textbf{80}, 1--28.

\bibitem[{{California Open Data Portal}(2022)}]{caopendata}
{California Open Data Portal} (2022) {California Open Data Portal}.
\newblock \url{https://data.ca.gov/dataset/covid-19-time-series-metrics-by-county-and-state1}.
\newblock [Online; accessed 19-Jan-2022].

\bibitem[{Capistr{\'a}n et~al.(2022)Capistr{\'a}n, Capella and Christen}]{capistran2022filtering}
Capistr{\'a}n, M.~A., Capella, A. and Christen, J.~A. (2022) Filtering and improved uncertainty quantification in the dynamic estimation of effective reproduction numbers.
\newblock \textit{Epidemics}, \textbf{40}, 100624.

\bibitem[{Champredon and Dushoff(2015)}]{champredon2015}
Champredon, D. and Dushoff, J. (2015) Intrinsic and realized generation intervals in infectious-disease transmission.
\newblock \textit{Proceedings of the Royal Society B: Biological Sciences}, \textbf{282}, 2015--2026.

\bibitem[{Champredon et~al.(2018)Champredon, Dushoff and Earn}]{champredon2018}
Champredon, D., Dushoff, J. and Earn, D. J.~D. (2018) Equivalence of the {Erlang-Distributed} {SEIR} {Epidemic Model} and the {Renewal Equation}.
\newblock \textit{SIAM Journal on Applied Mathematics}, \textbf{78}, 3258--3278.

\bibitem[{Corbella et~al.(2022)Corbella, Spencer and Roberts}]{corbella2022automatic}
Corbella, A., Spencer, S.~E. and Roberts, G.~O. (2022) {Automatic Zig-Zag sampling in practice}.
\newblock \textit{Statistics and Computing}, \textbf{32}, 107.

\bibitem[{Cori et~al.(2013)Cori, Ferguson, Fraser and Cauchemez}]{cori_new_2013}
Cori, A., Ferguson, N.~M., Fraser, C. and Cauchemez, S. (2013) A {New} {Framework} and {Software} to {Estimate} {Time}-{Varying} {Reproduction} {Numbers} {During} {Epidemics}.
\newblock \textit{American Journal of Epidemiology}, \textbf{178}, 1505--1512.

\bibitem[{Diggle et~al.(2002)Diggle, Diggle, Heagerty, Liang, Zeger et~al.}]{diggle2002analysis}
Diggle, P., Diggle, P.~J., Heagerty, P., Liang, K.-Y., Zeger, S. et~al. (2002) \textit{Analysis of longitudinal data}.
\newblock Oxford university press.

\bibitem[{Ferretti et~al.(2020)Ferretti, Ledda, Wymant, Zhao, Ledda, Abeler-D{\"o}rner, Kendall, Nurtay, Cheng, Ng, Lin, Hinch, Masel, Kilpatrick and Fraser}]{Ferretti2020}
Ferretti, L., Ledda, A., Wymant, C., Zhao, L., Ledda, V., Abeler-D{\"o}rner, L., Kendall, M., Nurtay, A., Cheng, H.-Y., Ng, T.-C., Lin, H.-H., Hinch, R., Masel, J., Kilpatrick, A.~M. and Fraser, C. (2020) The timing of {COVID-19} transmission.
\newblock \textit{medRxiv}.

\bibitem[{Fintzi et~al.(2022)Fintzi, Wakefield and Minin}]{stemr}
Fintzi, J., Wakefield, J. and Minin, V.~N. (2022) A linear noise approximation for stochastic epidemic models fit to partially observed incidence counts.
\newblock \textit{Biometrics}, \textbf{78}, 1530--1541.

\bibitem[{Flaxman et~al.(2020)Flaxman, Mishra and Gandy}]{flaxman_2020}
Flaxman, S., Mishra, S. and Gandy, A. e.~a. (2020) Estimating the effects of non-pharmaceutical interventions on {COVID}-19 in {Europe}.
\newblock \textit{Nature}, \textbf{584}, 257--261.
\newblock \urlprefix\url{https://doi.org/10.1038/s41586-020-2405-7}.

\bibitem[{Fraser(2007)}]{Fraser2007}
Fraser, C. (2007) Estimating {Individual} and {Household} {Reproduction} {Numbers} in an {Emerging Epidemic}.
\newblock \textit{PLOS ONE}, \textbf{2}, 1--12.

\bibitem[{Ganyani et~al.(2020)Ganyani, Kremer, Chen, Torneri, Faes, Wallinga and Hens}]{Ganyani2020}
Ganyani, T., Kremer, C., Chen, D., Torneri, A., Faes, C., Wallinga, J. and Hens, N. (2020) {Estimating} the generation interval for coronavirus disease {(COVID-19)} based on symptom onset data, {March} 2020.
\newblock \textit{Eurosurveillance}, \textbf{25}, 2000257.

\bibitem[{Gillespie(1977)}]{gillespie1977exact}
Gillespie, D.~T. (1977) {Exact Stochastic Simulation of Coupled Chemical Reactions}.
\newblock \textit{The Journal of Physical Chemistry}, \textbf{81}, 2340--2361.

\bibitem[{Gostic et~al.(2020)Gostic, McGough, Baskerville and Abbott}]{gostic2020}
Gostic, K.~M., McGough, L., Baskerville, E.~B. and Abbott, S. e.~a. (2020) Practical {Considerations} for {Measuring} the {Effective} {Reproductive} {Number}, {Rt}.
\newblock \textit{PLOS Computational Biology}, \textbf{16}, 1--21.

\bibitem[{Gronau et~al.(2020)Gronau, Singmann and Wagenmakers}]{bridgesampling}
Gronau, Q.~F., Singmann, H. and Wagenmakers, E.-J. (2020) {bridgesampling}: {An} {R} {Package} for {Estimating} {Normalizing} {Constants}.
\newblock \textit{Journal of Statistical Software}, \textbf{92}, 1--29.

\bibitem[{Hart et~al.(2022)Hart, Miller, Andrews, Waight, Maini, Funk and Thompson}]{hart2022}
Hart, W.~S., Miller, E., Andrews, N.~J., Waight, P., Maini, P.~K., Funk, S. and Thompson, R.~N. (2022) {Generation} time of the alpha and delta {SARS-CoV-2} variants: an epidemiological analysis.
\newblock \textit{The Lancet Infectious Diseases}, \textbf{22}, 603--610.

\bibitem[{Huisman et~al.(2022)Huisman, Scire, Angst, Li, Neher, Maathuis, Bonhoeffer and Stadler}]{huisman2022estimation}
Huisman, J.~S., Scire, J., Angst, D.~C., Li, J., Neher, R.~A., Maathuis, M.~H., Bonhoeffer, S. and Stadler, T. (2022) Estimation and worldwide monitoring of the effective reproductive number of sars-cov-2.
\newblock \textit{Elife}, \textbf{11}, e71345.

\bibitem[{Meng and Wong(1996)}]{meng1996simulating}
Meng, X.-L. and Wong, W.~H. (1996) Simulating {Ratios} of {Normalizing} {Constants} via a {Simple} {Identity}: a {Theoretical} {Exploration}.
\newblock \textit{Statistica Sinica}, 831--860.

\bibitem[{Mishra et~al.(2020)Mishra, Scott, Zhu, Ferguson, Bhatt, Flaxman and Gandy}]{Mishra2020}
Mishra, S., Scott, J., Zhu, H., Ferguson, N.~M., Bhatt, S., Flaxman, S. and Gandy, A. (2020) A {COVID-19} {Model} for {Local} {Authorities} of the {United} {Kingdom}.
\newblock \textit{medRxiv}.

\bibitem[{Nash et~al.(2022)Nash, Nouvellet and Cori}]{cori_review}
Nash, R.~K., Nouvellet, P. and Cori, A. (2022) {Real}-time estimation of the epidemic reproduction number: {Scoping} review of the applications and challenges.
\newblock \textit{PLOS Digital Health}, \textbf{1}, 1--17.

\bibitem[{Pakkanen et~al.(2022)Pakkanen, Miscouridou, Berah, Mishra, Mellan and Bhatt}]{pakkanen2022unifying}
Pakkanen, M.~S., Miscouridou, X., Berah, T., Mishra, S., Mellan, T.~A. and Bhatt, S. (2022) Unifying incidence and prevalence under a time-varying general branching process.
\newblock \textit{arXiv}.

\bibitem[{Parag(2021)}]{parag2021improved}
Parag, K.~V. (2021) Improved estimation of time-varying reproduction numbers at low case incidence and between epidemic waves.
\newblock \textit{PLoS Computational Biology}, \textbf{17}, e1009347.

\bibitem[{Park et~al.(2021)Park, Sun, Champredon, Li, Bolker, Earn, Weitz, Grenfell and Dushoff}]{Park2021}
Park, S.~W., Sun, K., Champredon, D., Li, M., Bolker, B.~M., Earn, D. J.~D., Weitz, J.~S., Grenfell, B.~T. and Dushoff, J. (2021) Forward-looking serial intervals correctly link epidemic growth to reproduction numbers.
\newblock \textit{Proceedings of the National Academy of Sciences}, \textbf{118}, e2011548118.

\bibitem[{Penn et~al.(2022)Penn, Laydon, Penn, Whittaker, Morgenstern, Ratmann, Mishra, Pakkanen, Donnelly and Bhatt}]{penn2022uncertainty}
Penn, M.~J., Laydon, D.~J., Penn, J., Whittaker, C., Morgenstern, C., Ratmann, O., Mishra, S., Pakkanen, M.~S., Donnelly, C.~A. and Bhatt, S. (2022) The uncertainty of infectious disease outbreaks is underestimated.
\newblock \textit{arXiv preprint arXiv:2210.14221}.

\bibitem[{{R Core Team}(2020)}]{R}
{R Core Team} (2020) \textit{R: A {Language} and {Environment} for {Statistical} {Computing}}.
\newblock R Foundation for Statistical Computing, Vienna, Austria.
\newblock \urlprefix\url{https://www.R-project.org/}.

\bibitem[{Scott et~al.(2021)Scott, Gandy, Mishra, Bhatt, Flaxman, Unwin and Ish-Horowicz}]{epidemia}
Scott, J.~A., Gandy, A., Mishra, S., Bhatt, S., Flaxman, S., Unwin, H. J.~T. and Ish-Horowicz, J. (2021) Epidemia: {An} {R} {Package} for {Semi}-{Mechanistic} {Bayesian} {Modelling} of {Infectious} {Diseases} using {Point} {Processes}.
\newblock \textit{arXiv}.

\bibitem[{Sender et~al.(2021)Sender, Bar-On, Park, Noor, Dushoff and Milo}]{Sender2021}
Sender, R., Bar-On, Y.~M., Park, S.~W., Noor, E., Dushoff, J. and Milo, R. (2021) The unmitigated profile of {COVID-19} infectiousness.
\newblock \textit{medRxiv}.
\newblock \urlprefix\url{https://www.medrxiv.org/content/early/2021/11/25/2021.11.17.21266051}.

\bibitem[{Sherratt et~al.(2021)Sherratt, Abbott, Meakin, Hellewell, Munday, Bosse, working group, Jit and Funk}]{sherratt2021exploring}
Sherratt, K., Abbott, S., Meakin, S.~R., Hellewell, J., Munday, J.~D., Bosse, N., working group, C. C.-., Jit, M. and Funk, S. (2021) Exploring surveillance data biases when estimating the reproduction number: with insights into subpopulation transmission of {COVID}-19 in {E}ngland.
\newblock \textit{Philosophical Transactions of the Royal Society B}, \textbf{376}, 20200283.

\bibitem[{{Stan Development Team}(2020)}]{rstan}
{Stan Development Team} (2020) {RStan}: the {R} interface to {Stan}.
\newblock \urlprefix\url{http://mc-stan.org/}.
\newblock R package version 2.21.2.

\bibitem[{Svensson(2007)}]{Svensson2007}
Svensson, A. (2007) A note on generation times in epidemic models.
\newblock \textit{Mathematical {Biosciences}}, \textbf{208}, 300--311.

\bibitem[{{Swiss National Covid-19 Science Task Force}(2020)}]{swiss_taskforce}
{Swiss National Covid-19 Science Task Force} (2020) Situation report: Reproductive number.
\newblock \urlprefix\url{https://ncs-tf.ch/en/situation-report}.
\newblock [Online; accessed 2020-09-17].

\bibitem[{Teh et~al.(2022)Teh, Elesedy, He, Hutchinson, Zaidi, Bhoopchand, Paquet, Tomasev, Read and Diggle}]{teh2022efficient}
Teh, Y.~W., Elesedy, B., He, B., Hutchinson, M., Zaidi, S., Bhoopchand, A., Paquet, U., Tomasev, N., Read, J. and Diggle, P.~J. (2022) Efficient {B}ayesian inference of instantaneous reproduction numbers at fine spatial scales, with an application to mapping and nowcasting the {COVID}-19 epidemic in {B}ritish local authorities.
\newblock \textit{Journal of the Royal Statistical Society Series A: Statistics in Society}, \textbf{185}, S65--S85.

\bibitem[{Thompson et~al.(2019)Thompson, Stockwin, van Gaalen, Polonsky, Kamvar, Demarsh, Dahlqwist, Li, Miguel, Jombart et~al.}]{thompson2019improved}
Thompson, R.~N., Stockwin, J.~E., van Gaalen, R.~D., Polonsky, J.~A., Kamvar, Z.~N., Demarsh, P.~A., Dahlqwist, E., Li, S., Miguel, E., Jombart, T. et~al. (2019) Improved inference of time-varying reproduction numbers during infectious disease outbreaks.
\newblock \textit{Epidemics}, \textbf{29}, 100356.

\bibitem[{Wallinga and Teunis(2004)}]{wallinga2004different}
Wallinga, J. and Teunis, P. (2004) Different epidemic curves for severe acute respiratory syndrome reveal similar impacts of control measures.
\newblock \textit{American Journal of Epidemiology}, \textbf{160}, 509--516.

\bibitem[{Wood(2017)}]{tp_splines}
Wood, S. (2017) \textit{Generalized {Additive} {Models}: {An} {Introduction} with {R} (2nd ed.)}.
\newblock Chapman and Hall/CRC.

\bibitem[{Xin et~al.(2022)Xin, Li, Wu, Li, Lau, Qin, Wang, Cowling, Tsang and Li}]{Xin2021}
Xin, H., Li, Y., Wu, P., Li, Z., Lau, E.~H., Qin, Y., Wang, L., Cowling, B.~J., Tsang, T.~K. and Li, Z. (2022) Estimating the {Latent} {Period} of {Coronavirus} {Disease} 2019 {(COVID-19)}.
\newblock \textit{Clinical Infectious Diseases}, \textbf{74}, 1678--1681.

\end{thebibliography}

\clearpage 

\appendix

\setcounter{page}{1}
\setcounter{table}{0}
\setcounter{equation}{0}
\setcounter{section}{0}
\setcounter{figure}{0}

\renewcommand\thefigure{\thesection\-\arabic{figure}}
\renewcommand\thetable{\thesection\-\arabic{table}}
\newpage
\section{Appendix}

\subsection{Methods}
\subsubsection{SEIR model used for simulation} \label{app:sim}
Here we describe in further detail the SEIR model used to simulate the data analyzed in this study. 
The SEIR model describes an infectious disease outbreak of a homogeneously mixing population, with the population divided into four compartments: susceptible, exposed (infected but not yet infectious), infectious, and removed.
The SEIR model is represented as a four dimensional continuous time Markov jump process, $\mathbf{G(t)} = (S(t), E(t), I(t), R=(t))$.
It can be defined in terms of rate parameters such that 
\begin{align*}
    P(\mathbf{G}(t + dt) &= (s - 1, e + 1, i, r)  \mid \mathbf{G}(t) = (s,e,i,r)) = \beta_{t}\times i \times s/N \times dt + o(dt),\\
    P(\mathbf{G}(t + dt) &= (s, e - 1, i + 1, r) \mid \mathbf{G}(t) = (s,e,i,r)) =  \gamma \times e \times dt + o(dt), \\ 
    P(\mathbf{G}(t + dt) &= (s, e, i - 1, r + 1) \mid \mathbf{G}(t) = (s,e,i,r)) = \nu \times i \times dt + o(dt). 
\end{align*}We use the well known Gillespie algorithm popularized in \citep{gillespie1977exact} to simulate from this model, as implemented in the \texttt{stemr} R package \citep{stemr}. 
Here $\gamma$ is the inverse of the mean latent period, and $\nu$ is the inverse of the mean infectious period. 
We describe the infectiousness of the disease through the time-varying transmission rate parameter $\beta_{t}$. 
With this model, the time-varying basic reproduction number, $R_{0,t}$, and effective reproduction number, $R_{t}$, are defined as
\begin{align*}
R_{0,t} = \frac{\beta_{t}}{\nu},\\
R_{t} =  R_{0,t} \times \frac{S(t)}{N}.
\end{align*}
By fixing the trajectory of $R_{0,t}$, we fix the trajectory of both $\beta_{t}$ but not $R_{t}$, because the susceptible population changes stochastically.
To simulate case data, we track cumulative incidence through a variable $C(t)$, which counts the transitions from the $E$ to the $I$ state.
Cases are then generated at a daily time-scale using the negative binomial model described in the methods section, changing the mean of the model so that, for day $t$:
\begin{align*}
    O_{t} \mid \mathbf{G}(t), \gamma, \nu, \boldsymbol{\beta}_{0:t}, \rho, \kappa, M_{t} &\sim \text{Neg-Binom}(\rho \times M_{t} \times (C(t) - C(t-1)), \kappa).
\end{align*}

\subsubsection{True $R_{t}$ curves of Scenario 1}
\begin{figure}[H]
    \centering
    \includegraphics[width = \textwidth]{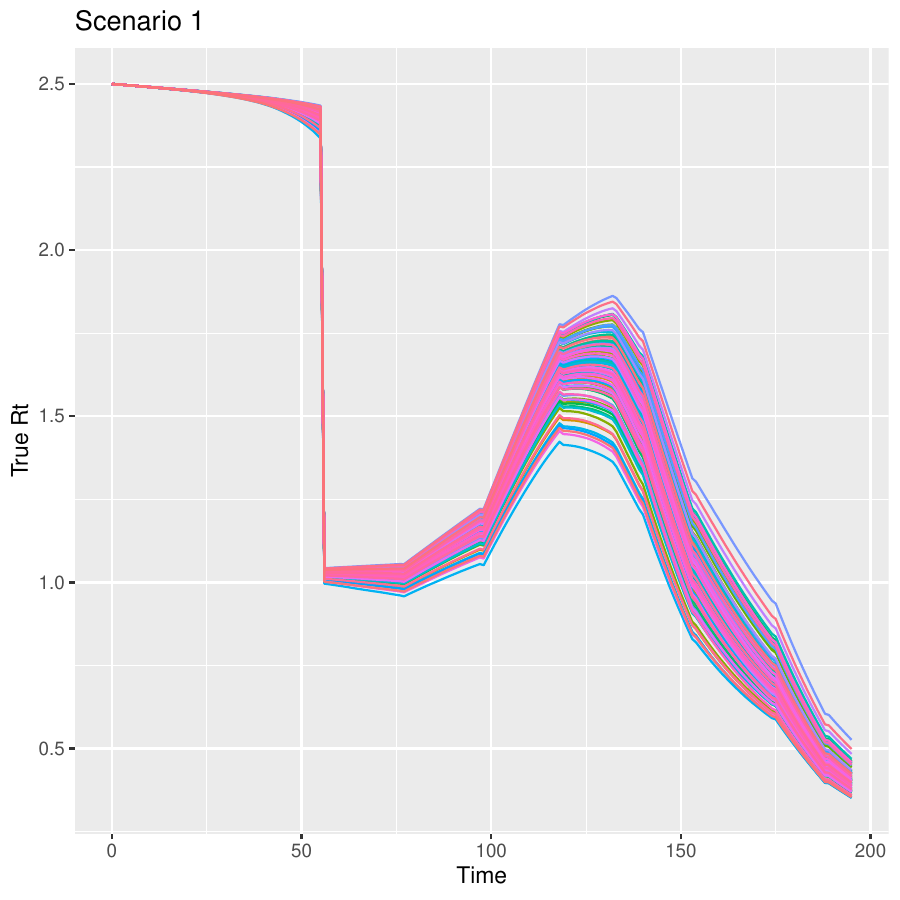}
    \caption{True $R_{t}$ curves for Scenario 1. The trajectory of $R_{0}$ is fixed, but because the number of susceptibles changes stochastically, each individual realization of the simulation has a slightly different $R_{t}$ trajectory.}
    \label{fig:s1_truert}
\end{figure}
\subsubsection{Rt-estim-normal model and parameters}
Below is the explicit model structure for the Rt-estim-normal model. 
 \begin{align*}
\lambda &\sim \text{exp}(\eta)  \quad \quad \text{Hyperprior for unobserved incidence}\\
I_{t} &\sim \text{exp}(\lambda) \quad \quad \text{Prior on unobserved incidence for t= -n, -n-1, \dots 0}\\
    \sigma & \sim \text{Truncated-Normal}(\mu_{\sigma}, \sigma_{\sigma}^{2}) \\
   \log{R_{1}} &\sim \text{Normal}(\mu_{r1}, \sigma_{r1}^{2})\quad \quad \text{Prior on $R_{1}$} \\
    \log{R_{t}}|\log{R_{t-1}} &\sim \text{Normal}(\log{R_{t-1}}, \sigma)  \quad \quad \text{Random Walk prior on $R_{t}$}\\
\psi &\sim \text{Normal}(\mu_{\psi},\sigma_{\psi}^2) \quad \quad \text{Prior on variance parameter for incidence} \\ 
I_{t}|I_{-n}, \dots, I_{t-1} &\sim \text{Normal}(R_{t}\sum_{s<t}I_{s}g_{t-s}, (R_{t}\sum_{s<t}I_{s}g_{t-s}*\psi)^{2}) \quad \quad \text{Model for incidence} \\
    \alpha &\sim \text{Normal}(\mu_{\alpha}, \sigma_{\alpha}^2)
\quad \quad \text{Prior on case detection rate} \\
y_{t} &= \alpha_{t}\sum_{s<t}I_{s}d_{t-s} \quad \quad \text{Mean of observed data model}\\
\frac{1}{\phi} & \sim \text{Normal}(\mu_{\phi}, \sigma_{\phi}^2) \quad \quad \text{Prior on dispersion parameter for observed data} \\
Y_{t} &\sim \text{Neg-Binom}(y_{t}, \phi) \quad \quad \text{Observed data model}\\
\end{align*}

The same priors were used for all simulations. They are described in Table \ref{tab:epidemia_sim_priors}. 
\begin{table}
    \caption{Priors used by the Rt-estim-normal method in the  simulation study.}
    \centering
    \begin{tabular}{cccc}
         Parameter &  Prior & Prior Median (95\% Interval)  \\
         \hline \\
         $\sigma$ & Truncated-normal(0, $0.1^{2}$) & 0.067 (0.0033, 0.26) \\
         $\lambda$ & Exponential(0.3) &  2.31 (0.08, 12.26) \\
         $\log{R_{1}}$ & Normal(0, $0.2^{2}$) & 0.00 (-0.39 0.39) \\ 
         $\psi$ & Normal(10, $2^{2}$) & 10 (6.11, 13.88) \\
         $\alpha$  &  Normal(0.02, $0.05^{2}$) & 0.02 (-0.08, 0.12) \\
         $\frac{1}{\phi}$ &  Normal(10, $5^{2}$) & 10 (0.23, 19.88) 
    \end{tabular}
    \label{tab:epidemia_sim_priors}
\end{table}

\subsubsection{Assessing model convergence in simulations}\label{convergence}
For Rt-estim-gamma, we assessed the minimum and maximum of the Rhat diagnostic, as well as the minimum and maximum effective sample size for each parameter. 
We considered maximum values of Rhat below 1.05 to indicate convergence, and considered effective sample size above 100 to be adequate. 
There were two instances in our original run of all simulations where the diagnostics were above these thresholds. 
For those specific simulations, we changed the seeds used to change the initial values of the MCMC, which led to convergence.
\subsubsection{Discretizing distributions}
The weights $g_{t-s}$ and $d_{t-s}$ used through the paper are discretized versions of continuous probability distributions. The number of discretized values to create was usually set to be the number of observed data points (occasionally with one additional value). For each value $u$ greater than 1, the discretized value was calculated as 
\begin{equation*}
    g_{u} = F(u + 0.5) - F(u - 0.5), u = 2, \dots
\end{equation*}
where $F(u)$ is the cumulative distribution function for the distribution being discretized. For $u=1$, in the case of the generation time distribution we used 
\begin{equation*}
    g_{1} = F(1.5),
\end{equation*}
but for the latent distribution, we used
\begin{align*}
    g_{0} &= F(0.5) \\
    g_{1} &= F(1.5) - F(0.5)
\end{align*}
in order to have a discretized value corresponding to 0. 
\subsubsection{Using Rt-estim-gamma with real data}
In practice, we have found it often necessary to provide reasonable initial values to start the Hamiltonian-Monte Carlo algorithm when applying the Rt-estim-gamma model to real data. 
Even so, running 4 chains, only three converged. We ran chains for 6000 iterations, discarding half as burn-in, and kept results only when the maximum Rhat value was calculated to be less than 1.05, with minimum bulk ESS and tail ESS above 100 as calculated using \texttt{rstan}. In all counties, the tail and bulk ESS for the estimates for the effective reproduction number had a minimum value of 1000. For the effective reproduction number, we first used \texttt{EpiEstim} to estimate the effective reproduction number, then used the median estimate from the posterior as the starting point for the effective reproduction number in Rt-estim-gamma. For incidence, we used the median of the overall case detection rate prior times the observed cases for the corresponding day. For all other model parameters, we used the mean of the prior distribution as the starting point. 

\subsubsection{Creating generation time distributions for delta and omicron variants}
We created generation time distributions for these variants by searching for parameters such that the new distributions had the appropriate new mean generation time, while preserving the standard deviation of the original distribution. We used a squared error loss function as a cost function, using the squared difference in a candidate distribution's mean vs the desired mean plus the squared difference in the candidate distribution's standard deviation vs the desired standard deviation. The estimates of the candidate distribution's mean and standard deviations were method of moment estimates from 100,000 samples generated in R. 
\subsection{Results}

\subsection{Posterior predictive distribution for three scenarios}
\begin{figure}[H]
    \centering
    \includegraphics[scale = 0.6]{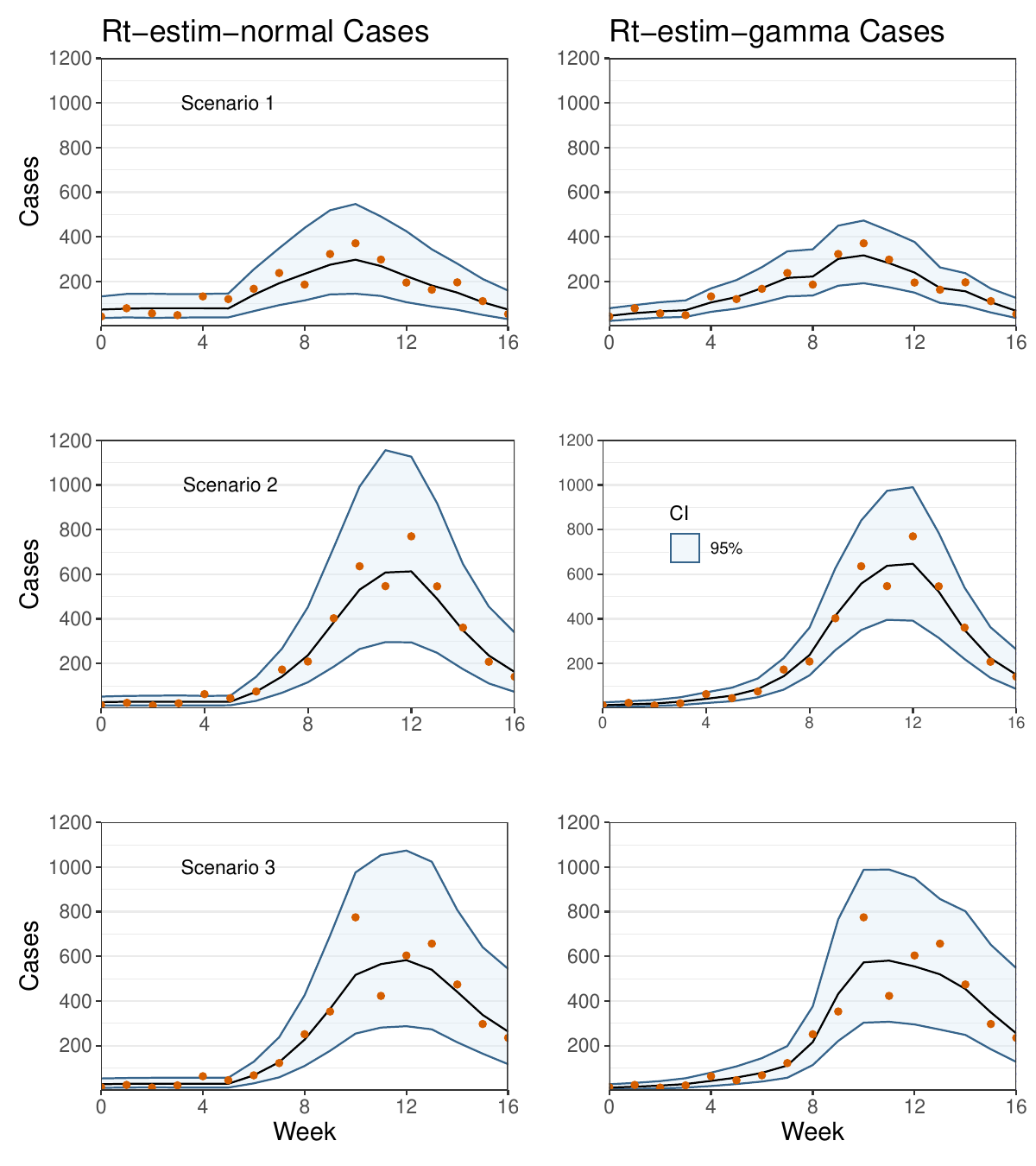}
    \caption{Posterior predictive distributions for reported cases using two different $R_{t}$ estimation methods for three simulated data sets under different testing scenarios. True incidence trajectories are colored in red, black lines represent median estimates from the posterior distribution, blue shaded ares are 95\% credible intervals.}
    \label{fig:sim_case}
\end{figure}

\subsubsection{Example negative-binomial spline posterior predictive}
\begin{figure}[H]
    \centering
    \includegraphics[width=\textwidth]{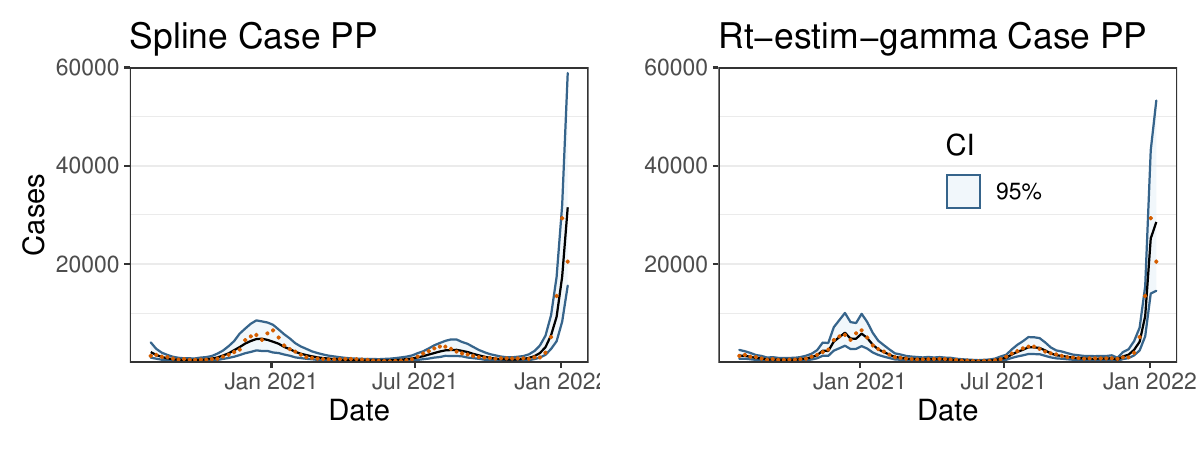}
    \caption{Posterior predictive plots of cases of SARS-CoV-2 in Alameda County, CA from August 20th 2020 through January 9th 2022. The left plot shows the posterior predictive distribution for a negative binomial spline, while the right plot shows the posterior predictive from the Rt-estim-gamma model. Black lines represent medians, blue bands 95\% credible intervals, and dots are observed cases.}
    \label{fig:spline_example}
\end{figure}
\subsubsection{Secondary simulation frequentist metrics}
\begin{figure}[H]
    \centering
     \includegraphics[width=\textwidth]{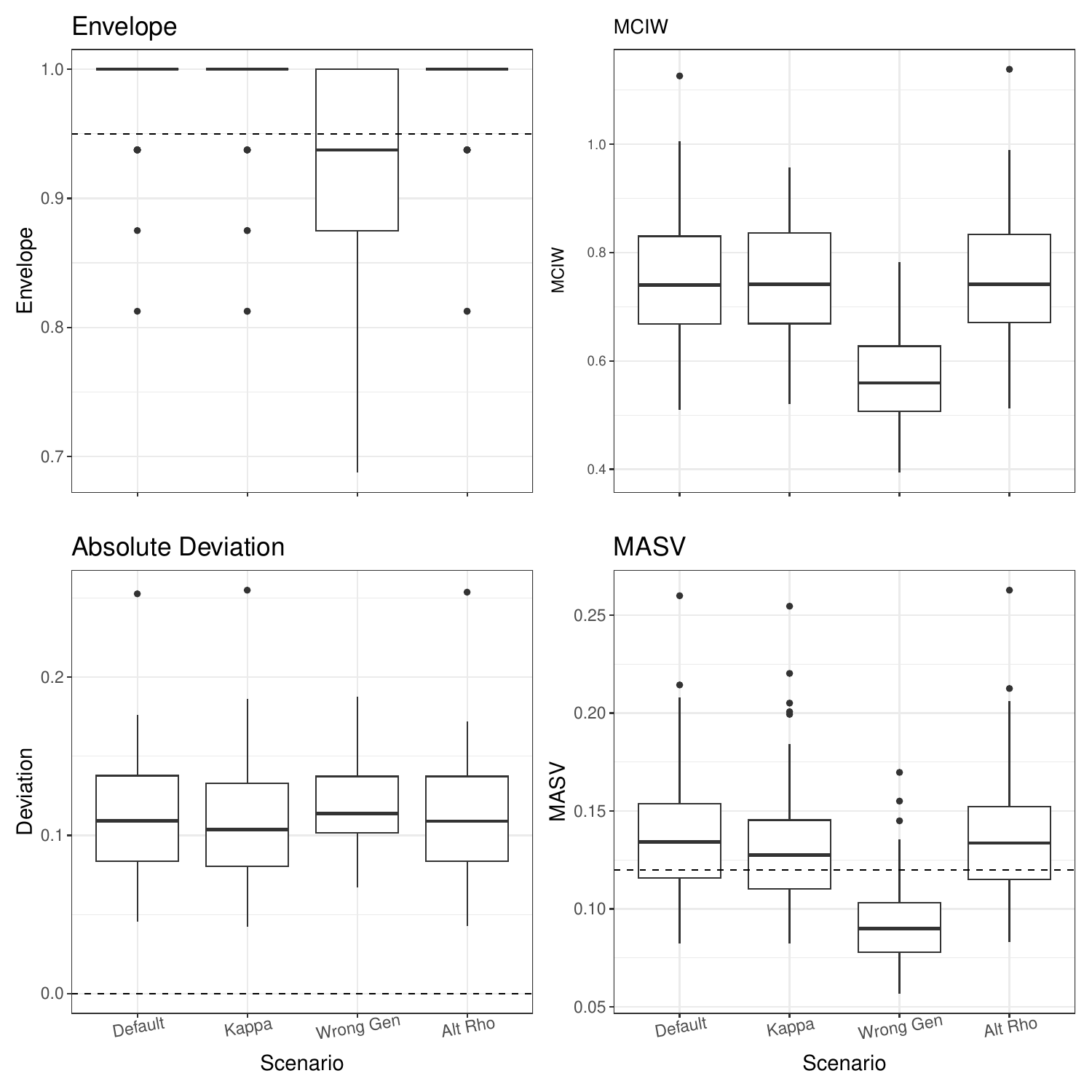}
    \caption{Frequentist metrics for Rt-estim-gamma applied to simulated epidemics from Scenario 3 with alternative model parameters. Default refers to the Scenario 3 results reported in the main text. Kappa refers to the choosing the priors for the overdispersion parameter of the case observation model by using a spline on the actual data being analyzed. Wrong gen refers to using a generation time distribution with rate parameters twice as large as the correct rate parameters. Alt rho refers to using a prior for $\rho$ based on the 25\% quantile of tests, rather than the 50\% quantile used in the main analysis. The dashed lines in the bottom row represent the true absolute standard deviation. Middle lines are medians across 100 simulations, hinges are upper and lower quartiles, whiskers are at most 1.5 times the interquartile range from the median.}
    \label{fig:app_freq_metric}
\end{figure}

\subsection{Using \texttt{EpiEstim} to estimate the effective reproduction number in CA}

\begin{figure}[H]
    \centering
    \includegraphics[width=\textwidth]{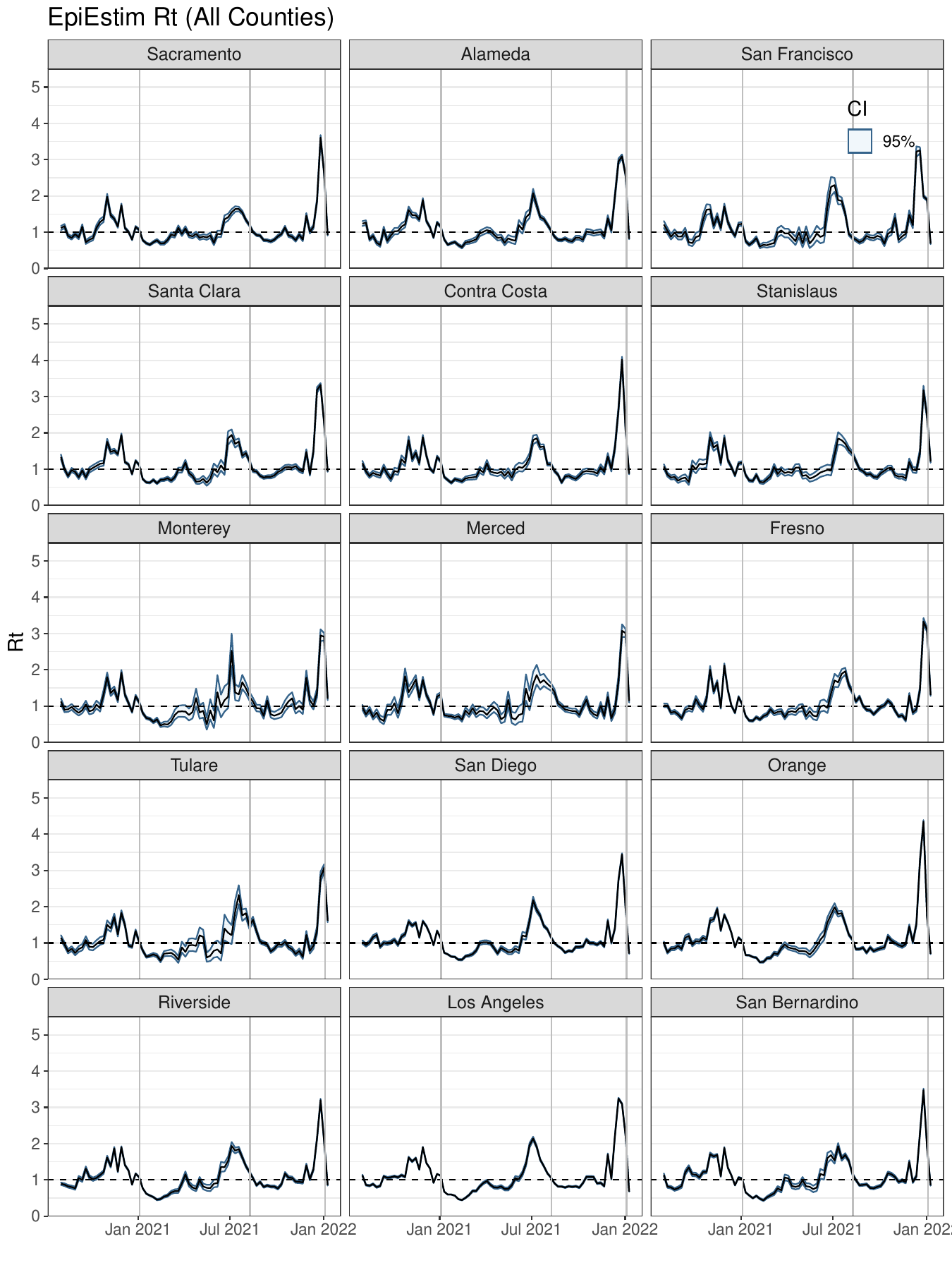}
    \caption{Estimates of the effective reproduction number in fifteen counties of California from August 2nd 2020 through January 9th 2022 using \texttt{EpiEstim}. Black lines represent medians, blue bars are 95\% credible intervals. }
    \label{fig:allcounties_epiestim}
\end{figure}

\subsection{Incidence posterior and case posterior predictive plots from Rt-estim-gamma for fifteen California counties}
\begin{figure}[H]
    \centering
    \includegraphics[scale = 0.55]{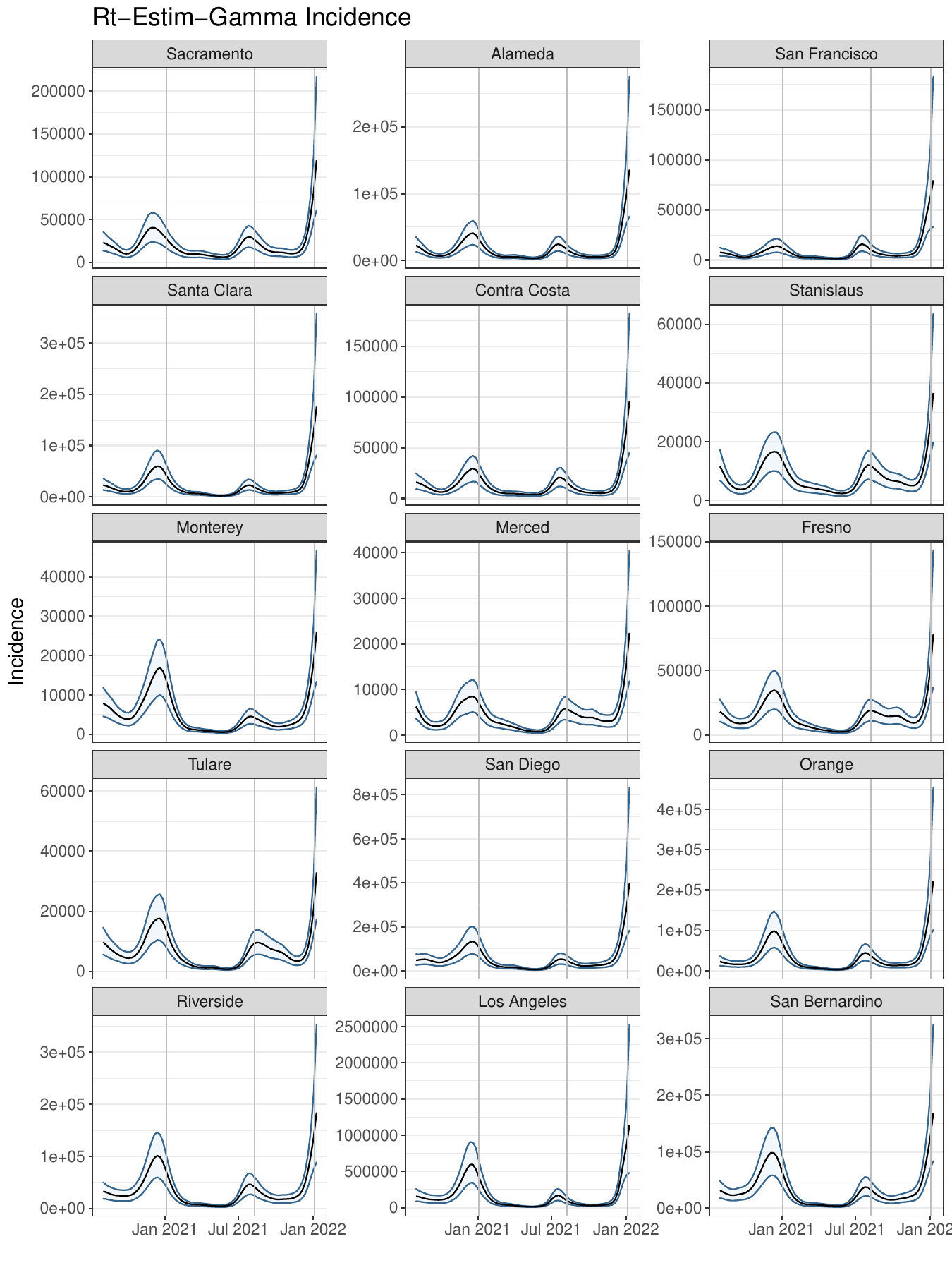}
    \caption{Estimates of incidence of SARS-CoV-2 from Rt-estim-gamma applied to fifteen counties in California, USA from August 2nd 2020 through January 15th 2022. Blue shaded regions are 95\% posterior credible intervals. Black lines are medians. Grey vertical lines mark the maximum statewide cases reported for the original winter 2020 wave, the delta-variant wave, and the omicron-variant wave.}
    \label{fig:allcounties_incid}
\end{figure}

\begin{figure}[H]
    \centering
    \includegraphics[width=\textwidth]{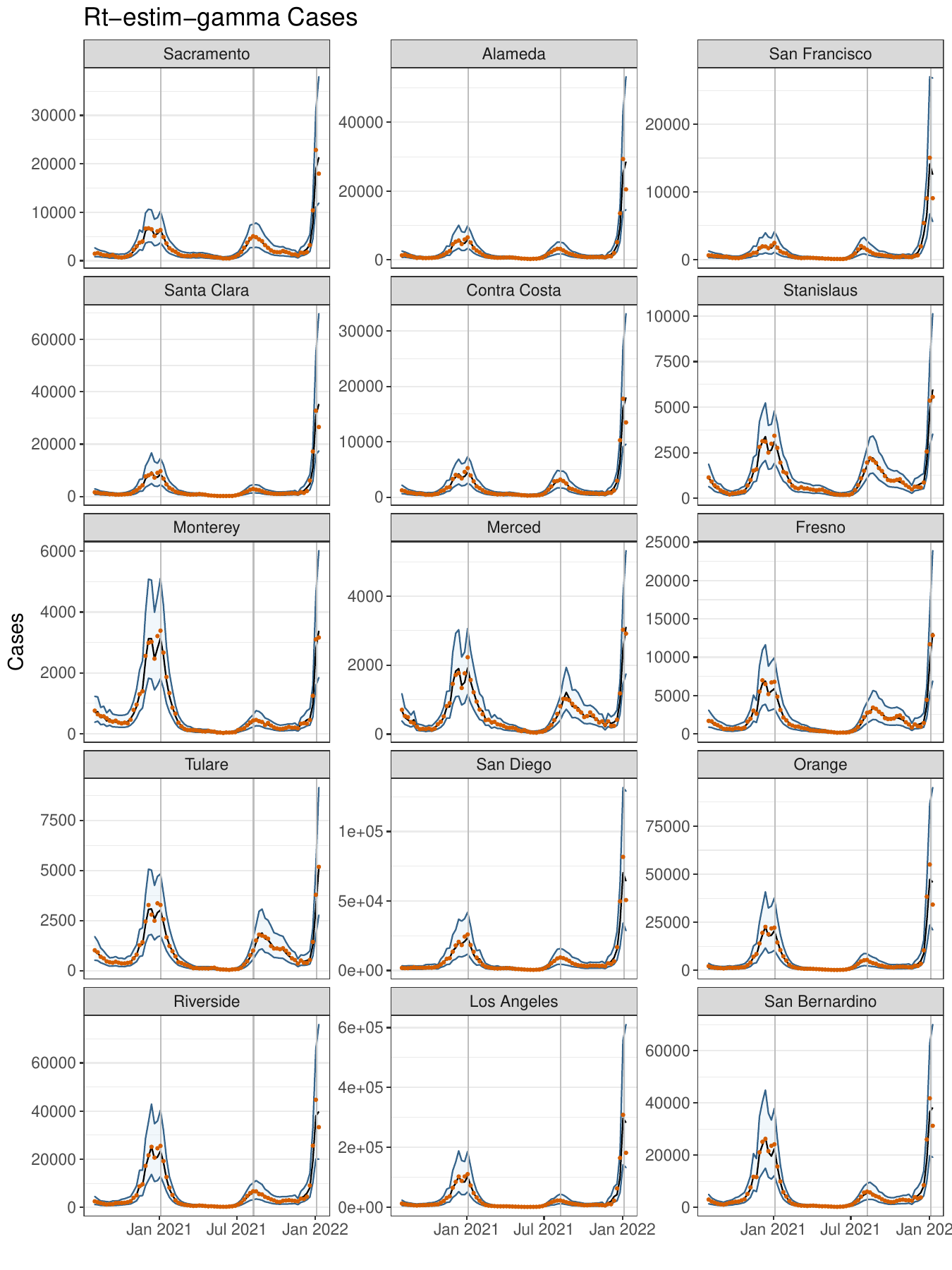}
    \caption{Posterior predictive estimates of reported cases of SARS-CoV-2 from Rt-estim-gamma applied to fifteen counties in California, USA from August 2nd 2020 through January 15th 2022. Blue shaded regions are 95\% posterior credible intervals. Black lines are medians. Grey vertical lines mark the maximum statewide cases reported for the original winter 2020 wave, the delta-variant wave, and the omicron-variant wave. Red dots are observed case counts.}
    \label{fig:allcounties_cases}
\end{figure}

\subsection{Comparing Rt-estim-normal and Rt-estim-gamma applied to fifteen California counties}
\begin{figure}[H]
    \centering
    \includegraphics[width=\textwidth]{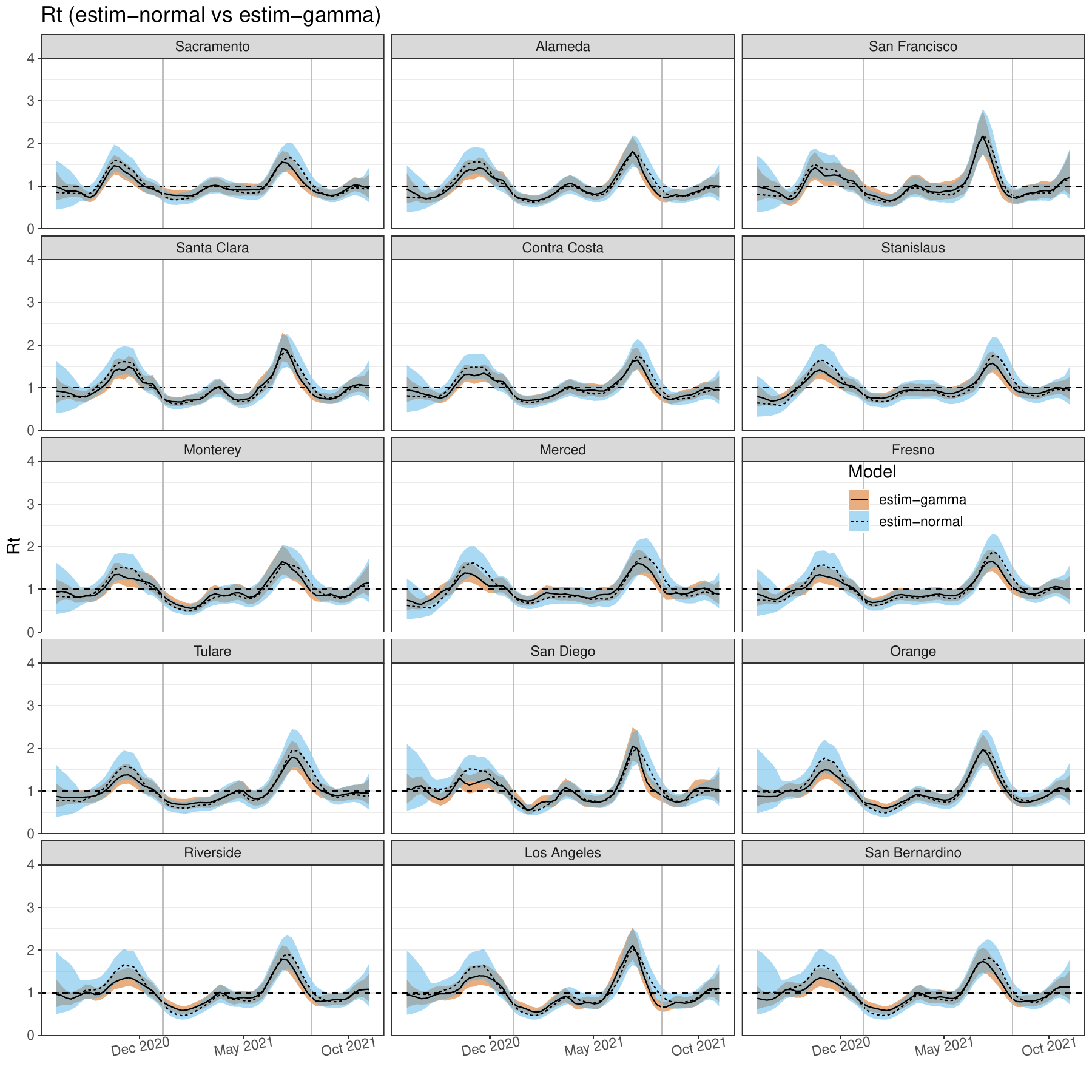}
    \caption{Estimates of the effective reproduction number of SARS-CoV-2 from Rt-estim-gamma and Rt-estim-normal applied to fifteen counties in California, USA from August 2nd 2020 through November 6th 2021. Blue  and brown shaded regions are 95\% posterior credible intervals. Black and dotted lines are medians. Grey vertical lines mark the maximum statewide cases reported for the original winter 2020 wave and the delta-variant wave. Blue shading and black lines come from estimates using Rt-estim-normal as opposed to brown shading with dotted lines, which denote estimates using Rt-estim-gamma.}
    \label{fig:ca_rt_estimnormal}
\end{figure}

\begin{figure}[H]
    \centering
    \includegraphics[width=\textwidth]{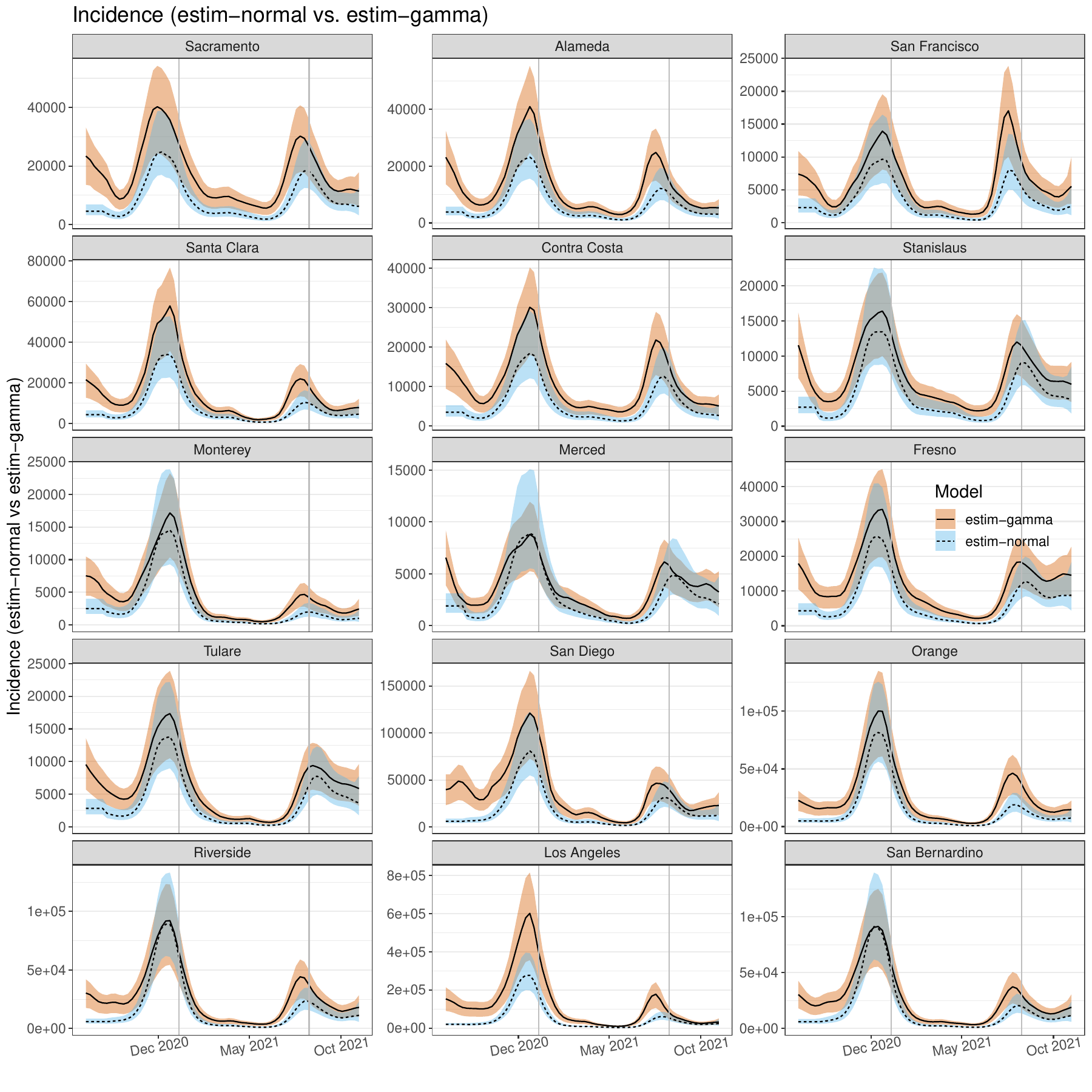}
    \caption{Estimates of incidence of SARS-CoV-2 from Rt-estim-gamma and Rt-estim-normal applied to fifteen counties in California, USA from August 2nd 2020 through November 6th 2021. Blue and brown shaded regions are 95\% posterior credible intervals. Black and dotted lines are medians. Grey vertical lines mark the maximum statewide cases reported for the original winter 2020 wave and the delta-variant wave. Blue shading and black lines come from estimates using Rt-estim-normal as opposed to brown shading with dotted lines, which denote estimates using Rt-estim-gamma.}
    \label{fig:ca_incid_estimnormal}
\end{figure}

\begin{figure}[H]
    \centering
    \includegraphics[width=\textwidth]{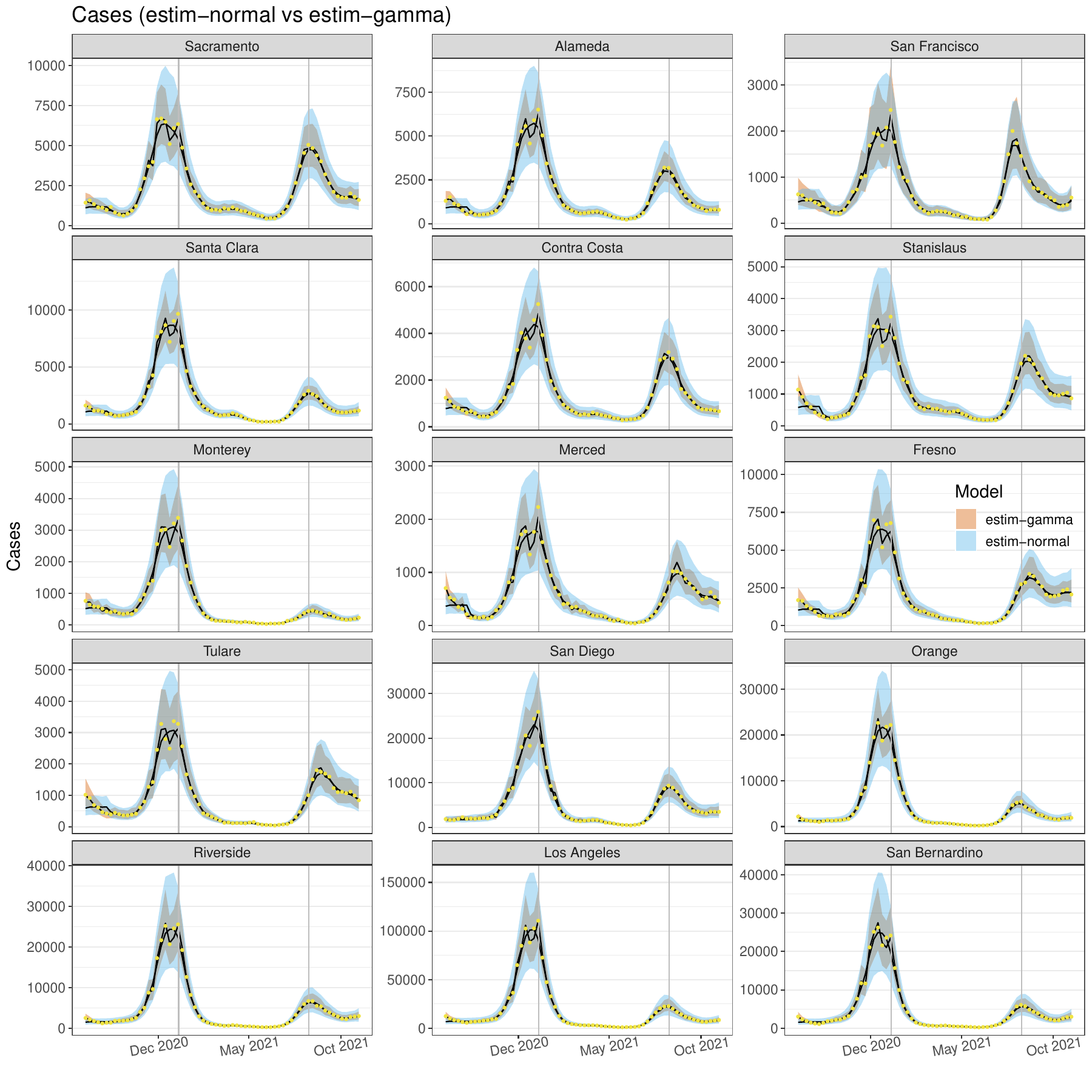}
    \caption{Estimates of observed cases of SARS-CoV-2 from Rt-estim-gamma and Rt-estim-normal applied to fifteen counties in California, USA from August 2nd 2020 through November 6th 2021. Blue and brown shaded regions are 95\% posterior predictive intervals intervals. Black and dotted lines are medians. Grey vertical lines mark the maximum statewide cases reported for the original winter 2020 wave and the delta-variant wave. Blue shading and black lines come from estimates using Rt-estim-normal as opposed to brown shading with dotted lines, which denote estimates using Rt-estim-gamma. Yellow dots are observed cases}
    \label{fig:ca_cases_estimnormal}
\end{figure}
\subsection{Comparing Rt-estim-gamma applied to fifteen California counties using a mean generation time of 5.5 days vs 9.7 days}
\begin{figure}[H]
    \centering
    \includegraphics[width=\textwidth]{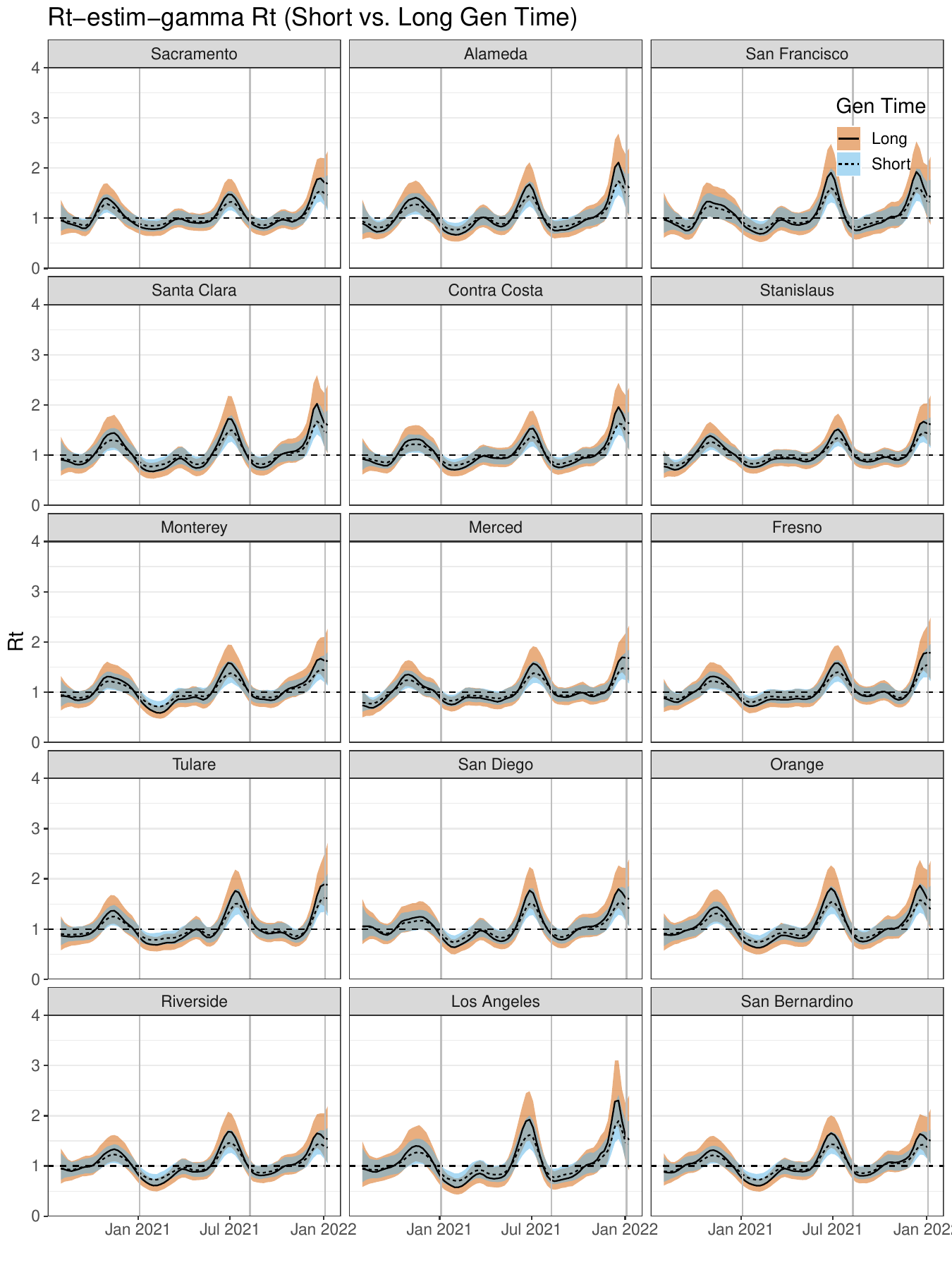}
    \caption{Estimates of the effective reproduction number of SARS-CoV-2 from Rt-estim-gamma applied to fifteen counties in California, USA from August 2nd 2020 through January 15th 2022. Blue and brown shaded regions are 95\% posterior credible intervals. Black and dotted lines are medians. Grey vertical lines mark the maximum statewide cases reported for the original winter 2020 wave, the delta-variant wave, and the omicron-variant wave. Blue shading and dotted lines come from estimates using mean generation time of 5.5 days as opposed to brown shading with dotted lines, which denote estimates using 9.7 days.}
    \label{fig:allcounties_rt_shortgen}
\end{figure}

\begin{figure}[H]
    \centering
    \includegraphics[width=\textwidth]{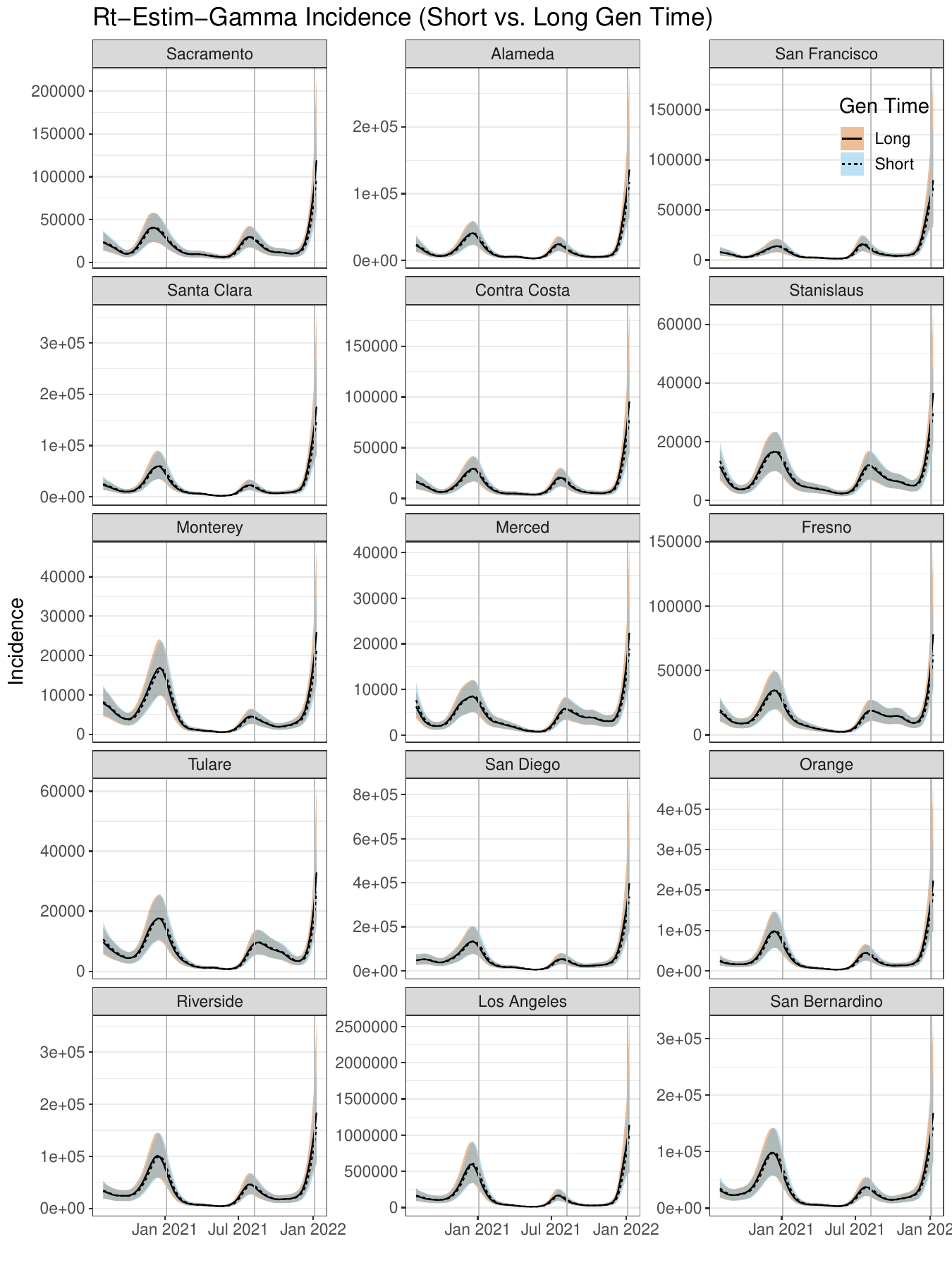}
    \caption{Estimates of incidence of SARS-CoV-2 from Rt-estim-gamma applied to fifteen counties in California, USA from August 2nd 2020 through January 15th 2022. Blue  and brown shaded regions are 95\% posterior credible intervals. Black and dotted lines are medians. Grey vertical lines mark the maximum statewide cases reported for the original winter 2020 wave, the delta-variant wave, and the omicron-variant wave. Blue shading and dotted lines come from estimates using mean generation time of 5.5 days as opposed to brown shading with dotted lines, which denote estimates using 9.7 days.}
    \label{fig:allcounties_incid_shortgen}
\end{figure}

\begin{figure}[H]
    \centering
    \includegraphics[width=\textwidth]{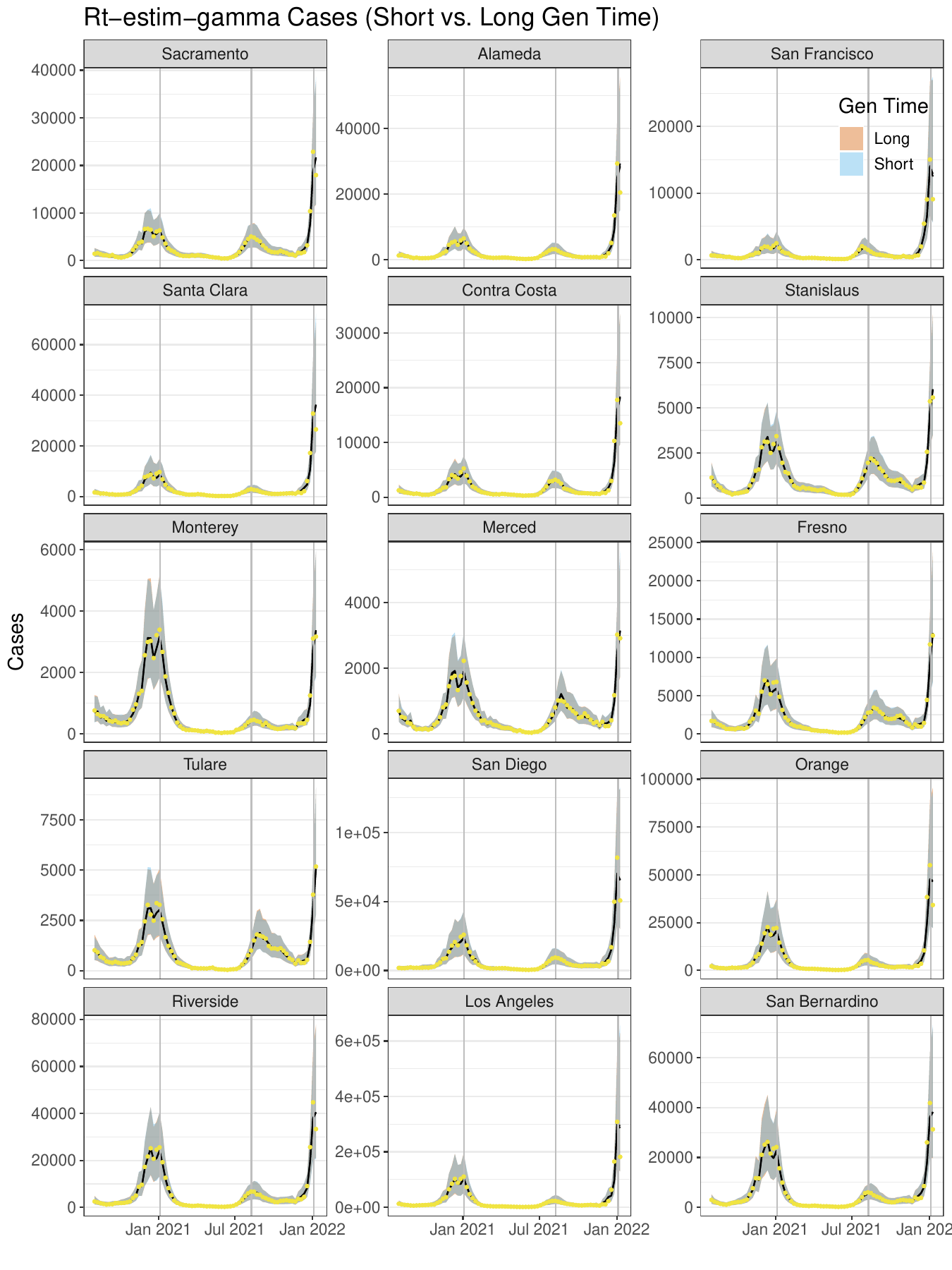}
    \caption{Posterior predictive estimates of reported cases of SARS-CoV-2 from Rt-estim-gamma applied to fifteen counties in California, USA from August 2nd 2020 through January 15th 2022. Blue and brown shaded regions are 95\% posterior credible intervals, grey represents overlap between the two estimates. Grey vertical lines mark the maximum statewide cases reported for the original winter 2020 wave, the delta-variant wave, and the omicron-variant wave. Yellow dots are observed case counts. Blue shading come from estimates using mean generation time of 5.5 days as opposed to brown shading which denote estimates using 9.7 days.}
    \label{fig:allcounties_cases_shortgen}
\end{figure}

\subsection{Prior and Posterior of Fixed Parameters}
\begin{figure}[H]
    \centering
    \includegraphics[width=\textwidth]{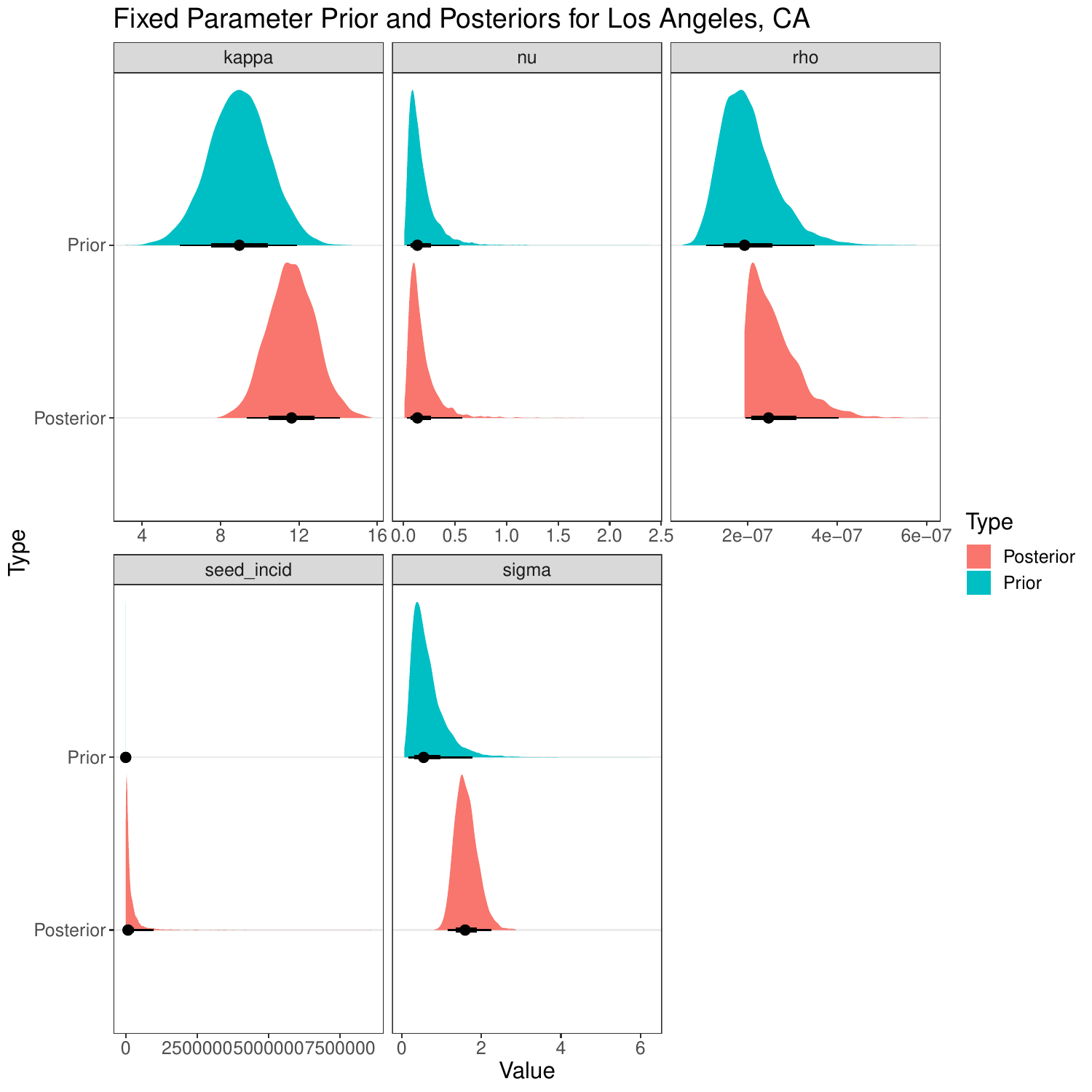}
    \caption{Priors and posteriors for fixed parameters from Rt-estim-gamma fit to Los Angeles, CA data using the Sender generation time. 
    The \text{seed\_incid} refers to the first unobserved incidence used by the model.}
    \label{fig:la_prior_post}
\end{figure}
\end{document}